\documentclass{aa}  
\usepackage{graphicx}
\usepackage{txfonts}
\usepackage[colorlinks=true,citecolor=blue]{hyperref}
\usepackage{multirow}
\usepackage{academicons}
\usepackage{amsmath,amstext}
\usepackage[T1]{fontenc}
\usepackage{color}

\DeclareRobustCommand{\ion}[2]{%
\relax\ifmmode
\ifx\testbx\f@series
{\mathbf{#1\,\mathsc{#2}}}\else
{\mathrm{#1\,\mathsc{#2}}}\fi
\else\textup{#1\,{\mdseries\textsc{#2}}}%
\fi}

\def\OIIIHb{[{\ion{O}{iii}}]/H$\beta$}
\def\OIII5007Hb{[{\ion{O}{iii}}] $\lambda5007$/H$\beta$}
\def\ratioR23{([\ion{O}{ii}] $\lambda$3727 +[\ion{O}{iii}] $\lambda\lambda$4959,5007)/H$\beta$}
\def\R23{${\rm R}_{23}$}
\def\dS23{${\rm S}_{23}$}

\def\ratioS23{([\ion{S}{2}] $\lambda \lambda$6717,31 +[\ion{S}{3}] $\lambda\lambda$9069,9532)/H$\beta$}
\def\NIIHa{[\ion{N}{ii}]/H$\alpha$}

\def\L60{L$_{60}$}

\def\Rv{$R_V$}
\def\EBV{E($B-V$)}
\def\B26{${{\rm B}_{26}}$}
\def\I26{${\rm I_{B_{26}}}$}

\newcommand{\orcid}[1]{\href{https://orcid.org/#1}{\includegraphics[width=10pt]{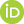}}}

\def\OIIIu{[{\ion{O}{iii}}] $\lambda$4959}

\def\OIIId{[{\ion{O}{iii}}] $\lambda$5007}

\def\OIIIt{[{\ion{O}{iii}}] $\lambda$4363}
\def\OIII{[{\ion{O}{iii}}]}

\def\NIId{[{\ion{N}{ii}}] $\lambda$6583}

\newcommand{\Had}{\rm{H}$\alpha$}

\newcommand{\Hbd}{\rm{H}$\beta$}

\newcommand{\salttwo}{{\sc{SALT2}}}

\newcommand{\lc}{light-curve}

\begin{document} 

\title{Aperture-corrected spectroscopic type Ia supernova \\ host galaxy properties}
\titlerunning{Aperture-corrected SNIa host properties}

\author{Llu\'is Galbany\inst{1,2}\fnmsep\thanks{\email{lgalbany@ice.csic.es}}\orcid{0000-0002-1296-6887},
Mat Smith\inst{3},
Salvador Duarte Puertas\inst{4,5},
Santiago Gonz\'alez-Gait\'an\inst{6},
Ismael Pessa\inst{7},\\
Masao Sako\inst{8}\orcid{0000-0003-2764-7093},
Jorge Iglesias-P\'aramo\inst{5}, 
A. R. L\'opez-S\'anchez\inst{9,10,11,12},
Mercedes Moll\'a\inst{13},
Jos\'e M. V\'ilchez\inst{5}.
}
\authorrunning{Galbany et al.}

\institute{
Institute of Space Sciences (ICE, CSIC), Campus UAB, Carrer de Can Magrans, s/n, E-08193 Barcelona, Spain.
\and Institut d’Estudis Espacials de Catalunya (IEEC), E-08034 Barcelona, Spain.
\and Univ Lyon, Univ Claude Bernard Lyon 1, CNRS, IP2I Lyon / IN2P3, IMR 5822, F-69622, Villeurbanne, France.
\and D\'epartement de Physique, de G\'enie Physique et d’Optique, Universit\'e Laval, and Centre de Recherche en Astrophysique du Qu\'ebec (CRAQ), Qu\'ebec, QC, G1V 0A6, Canada
\and Instituto de Astrof\'isica de Andaluc\'ia - CSIC, Glorieta de la Astronom\'ia s.n., 18008 Granada, Spain.
\and CENTRA/COSTAR, Instituto Superior T\'ecnico, Universidade de Lisboa, Av. Rovisco Pais 1, 1049-001 Lisboa, Portugal.
\and Max-Planck-Institute for Astronomy, K\"onigstuhl 17, D-69117 Heidelberg, Germany
\and Department of Physics and Astronomy, University of Pennsylvania, Philadelphia, PA 19104, USA.
\and Australian Astronomical Optics, Macquarie University, 105 Delhi Rd, North Ryde, NSW 2113, Australia.
\and Department of Physics and Astronomy, Macquarie University, NSW 2109, Australia.
\and Macquarie University Research Centre for Astronomy, Astrophysics \& Astrophotonics, Sydney, NSW 2109, Australia.
\and ARC Centre of Excellence for All Sky Astrophysics in 3 Dimensions (ASTRO-3D), Australia.
\and Dpto. de Investigaci\'on B\'asica, CIEMAT, Avda. Complutense 40, E-28040 Madrid, Spain.
}

\date{Received June 18, 2021; accepted hopefully soon.}

\abstract
{We use type Ia supernovae (SNe Ia) data obtained by the Sloan Digital Sky Survey-II Supernova Survey (SDSS-II/SNe) in combination with the publicly available SDSS DR16 fiber spectroscopy of their host galaxies to correlate SNe Ia light-curve parameters and Hubble residuals to several host galaxy properties. 
Fixed-aperture fiber spectroscopy suffers from aperture effects: the fraction of the galaxy covered by the fiber varies depending on its projected size on the sky, thus measured properties are not representative of the whole galaxy. 
The advent of Integral Field Spectroscopy has provided a way for correcting the missing light, by studying how these galaxy parameters change with the aperture size.
Here we study how the standard SN host galaxy relations change once global host galaxy parameters are corrected for aperture effects.
We recover previous trends on SN Hubble residuals with host galaxy properties, but we find that discarding objects with poor fiber coverage instead of correcting for aperture loss introduces biases in the sample that affect SN host galaxy relations.
The net effect of applying the  commonly used $g$-band fraction criterion is discarding intrinsically faint \mbox{SNe~Ia} in high-mass galaxies, thus artificially increasing the height of the mass step by 0.02 mag and its significance. 
Current and next generation of fixed-aperture fiber spectroscopy surveys, such as DES, DESI or TiDES in 4MOST, that aim at study SN and galaxy correlations must consider, and correct for, these effects. 
}

\keywords{supernovae: general --- supernovae: distances --- supernovae: host galaxies}

\maketitle


\section{Introduction}

The discovery of the accelerated expansion of the Universe was possible thanks to the use of type Ia supernovae (SNe Ia) as distance indicators \citep{1998AJ....116.1009R, 1999ApJ...517..565P}. Given their extreme brightness, SNe Ia can be detected even in very distant galaxies (z$\gtrsim1$) from the ground with mid-size ($\sim$4m) telescopes. At optical wavelengths, they show a scatter of $\sim$2 mag in their peak brightness, but it can be standardized by accounting for empirical relations with the brightness decay rate (or light-curve width; \citealt{1993ApJ...413L.105P}) and color at peak \citep{,1996ApJ...473...88R}. Once standardized, the precision in the distance measurements is reduced down to 5-7\% \citep{2014A&A...568A..22B,2019ApJ...874..150B}.

Over the years, the accumulation of new observations have weighted down statistical errors and the focus is now in the reduction of the systematic uncertainties. Such systematics can be reduced by improving the cadence of the observed light-curves \citep{2020ApJ...890...60R}, the photometric calibration of the fitting and minimization routines, or by modifying the standardization method either adding new parameters to the widely used Tripp expression \citep{1998A&A...331..815T} or exploring other (non-linear) approaches \citep{2015ApJ...813..137R,2020arXiv200807538M}. There is also evidence that near-infrared (NIR) observations, which are naturally less affected by reddening, present very low scatter on their peak magnitudes even with no corrections, which turns them into almost natural standard candles \citep{2018A&A...609A..72D,2018ApJ...869...56B}. Also other cosmological effects that affect measurements at high redshift (e.g. weak lensing; \citealt{2014ApJ...780...24S}) need to be accounted for in reducing the systematic uncertainty budget. 

One of these sources of systematics is the different environments where SNe Ia occur. Given that the characterization of the near local vicinity is problematic for farther objects (see for instance \citealt{2013A&A...560A..66R,2020A&A...644A.176R,2018ApJ...867..108J,2021MNRAS.501.4861K}), these efforts have been focused on measurements of the global properties of SN host galaxies, such as morphology, mass, age, star-formation rate (SFR), or metallicity.
Several correlations between SN properties and their environment have been reported so far: \cite{1996AJ....112.2391H} first found a correlation between SN Ia stretch and the Hubble type of its host, late-type galaxies hosting observationally brighter events compared to early-type hosts. This morphological classification is related to other galactic properties, such as the SFR, age, or mass. \cite{2006ApJ...648..868S} showed that high-stretch events were found in systems with ongoing star-formation, corresponding to the late-type galaxies. In terms of age of the stellar populations, it was also found that galaxies having younger mean ages host observationally brighter SNe \citep{1995AJ....109....1H, 2000AJ....120.1479H}. These correlations have also been confirmed in more recent works with larger SN samples by 
\cite{2005ApJ...634..210G}, 
\cite{2011ApJ...740...92G}, \cite{2013ApJ...770..108C}, \cite{2019ApJ...874...32R}, and \cite{2020MNRAS.494.4426S}.
The total mass of the galaxy has also been correlated to both the observed and the standardized SN brightness \citep{2010ApJ...715..743K, 2010MNRAS.406..782S}, providing the base for the introduction of a new term in the standardization equation that accounts for a host environmental parameter ($\gamma$ mass term; \citealt{2010ApJ...722..566L}). More recently, further efforts have been focused on studying this dependence at NIR wavelenghts \citep{2020ApJ...901..143U,2020arXiv200613803P}, and its relation with dust and the extinction law \citep{2021ApJ...909...26B,2021MNRAS.508.4656G}.

SN host galaxy parameters are primarily obtained from photometric magnitudes. Obtaining imaging of SN host galaxies is relatively easy and it may be available from observations of the same rolling surveys either before the SN or after when the SN is already undetected, or even in archival images. However, host galaxy parameters measured from photometry are less precise than when measured from spectroscopy, since the whole spectral energy distribution (SED) and properties of the galaxy are inferred from only a few photometric points. Although spectroscopy provides more precise information, it also presents some difficulties. Obtaining spectroscopy of SN host galaxies is difficult and far from common in SN surveys.

Problematically, the full extent of the galaxy is not usually covered by the fiber/slit, and galactic parameters measured from the spectra are representative only of the fraction of the area covered by the fiber/slit. To overcome this problem, one common approach followed in the literature is to scale the spectrum to match the integrated light of the galaxy. Although it helps with extensive properties (such as the stellar mass or the SFR, it still retains the assumption of extrapolating the properties of the area covered within the fiber/slit to the other regions of the galaxy (usually the outskirts), where intensive properties are known to change (such as metallicty, age, extinction, etc... in radial gradients, or structurally, e.g. arm-interarm; \citealt{2015A&A...573A.105S,2020MNRAS.492.4149S}). Another approach would be to use the property measured from the spectrum as a proxy for the central parameter, and use gradients to infer these properties at larger galactocentric distances. However, by far the best approach would be to obtain spectra of the whole extent of the galaxy through integral field spectroscopy (IFS), and sum them up to get real integrated properties. Although this is the most time expensive approach, new instrumentation and surveys are starting to build up the necessary samples (e.g. \citealt{2016MNRAS.455.4087G,2018ApJ...855..107G}).

Here we present a study of how correlations found between SN Ia parameters and fiber-spectroscopy host galaxy properties change when proper aperture corrections are applied. For that, we use the Sloan Digital Sky Survey-II Supernova Survey (SDSS-II/SNe) sample and publicly available spectroscopy of their host galaxies obtained with the SDSS and the SDSS's Baryon Oscillation Spectroscopic Survey (BOSS) fiber spectrographs in combination with aperture corrections derived using IFS of nearby galaxies from the CALIFA Survey \citep{2013A&A...553L...7I,2016ApJ...826...71I}. In this work we focus on gas-phase emission line parameters, namely SFR, oxygen abundance, extinction, and H$\alpha$ equivalent width, which are the parameters for which aperture corrections have been produced so far. Therefore, our galaxy sample will consist only of star-forming galaxies. Our approach is compared to the widely used criterion of discarding galaxies whose $g$-band light fraction covered by the fiber is lower than 20\% of the total galaxy light.

The paper is structured as follows. In \S\ref{sec:SN} we present the selection of the initial SN and host galaxy sample. In \S\ref{sec:HG} we describe all the procedures performed to obtain the needed host galaxy parameters. The effect of aperture corrections on our measurements, and the biases introduced by other approaches is in \S\ref{sec:eff}. In \S\ref{sec:sifto} we present the SN light-curve parameters and Hubble residuals, and in the following section (\S\ref{sec:results}) we present and discuss our results. Finally, we conclude in \S\ref{sec:disc}. Throughout the paper a flat $\Lambda$CDM cosmology with $\Omega_M=0.27$ and $\Omega_\Lambda=0.73$, and $H_0$=70.8 km~s$^{-1}$~Mpc$^{-1}$, is assumed.


\section{Supernova and host galaxy sample} \label{sec:SN}

\subsection{SDSS-II SN Survey}

We use the SN Ia sample provided by the SDSS-II/SNe in their Data Release (DR, \citealt{2018PASP..130f4002S}), consisting of 1364 SNe Ia either confirmed spectroscopically (540 spec-Ia) or photometrically identified based on a Bayesian LC fitting using the spectroscopic redshift of their host galaxy (824 photo-Ia, \citealt{2014AJ....147...75O,2018PASP..130f4002S}), with lightcurves in five ($ugriz$; \citealt{1996AJ....111.1748F}) bands. Observations were performed with the dedicated SDSS 2.5m telescope at Apache Point Observatory \citep{Gunn:1998p1708, Gunn:2006p159} during the three Fall seasons of operation from 2005 to 2007. Additionally, we included the 16 SNe Ia discovered during the 2004 pilot season of the survey. All SNe were all located in Stripe 82, a 300 deg$^2$ region along the Celestial Equator in the Southern Galactic hemisphere \citep{Stoughton:2002p547}. More details on the survey, photometry, and cosmological results can be found in \cite{2008AJ....135..338F}, \cite{2008AJ....136.2306H}, \cite{2009ApJS..185...32K}, and \cite{2018PASP..130f4002S}.

The SDSS-II/SNe DR provided a list of object ID corresponding to the associated host galaxies in the SDSS database. However, we performed our own SN-galaxy matching: we have produced 40"$\times$40" finding charts for all SNe in the sample from the SDSS DR16 \citep{2019arXiv191202905A}, and visually inspected for publicly available host galaxy spectra in DR16. Most SNe Ia were matched to a host galaxy following these two criteria: (i) the galaxy had to be within an angular separation of 20 arcsec from the SN \citep{2012ApJ...755..125G}, and (ii) the SN and the galaxy redshifts had to be equivalent within 3\%. 

After the visual inspection described above, our sample consists of optical spectra of the 1066 (352 spec-Ia and 714 photo-Ia) remaining host galaxies from SDSS DR16. Overall, we find a difference in the determined host galaxy for 24 objects (2\%) compared to \cite{2018PASP..130f4002S}. In this analysis we only considered spectra obtained with the SDSS and BOSS spectrographs, which is available in the SDSS DR16, and data obtained from other facilities (comprising 55 events) is excluded. More details on the host galaxy matching can be found in Appendix \ref{sec:sdss}. In addition, we also obtained multi-band photometric parameters of these galaxies needed for the rescaling of the spectrum to match the photometry (Section \S\ref{sec:resc}). 
The selection of the sample described above has been also used in the accompanying paper by \cite{2018MNRAS.476..307M}.

The spectroscopically confirmed SDSS-II/SNe sample (spec-Ia) is considered to be complete up to z=0.25 since the detection efficiency remained high up to this redshift ($\sim$95\%, \citealp{2008ApJ...682..262D, 2010ApJ...713.1026D, 2012ApJ...755...61S}). However, the addition of the photo-Ia sample has the effect of correcting for the missing objects at higher redshifts  increasing the completeness from 65\% to 85\% at z$\sim$0.4 \citep{2012ApJ...755...61S, 2013ApJ...763...88C}.


\section{Host galaxy characterization} \label{sec:HG}

To extract spectroscopic host galaxy parameters and study the effect of aperture corrections, we will follow three different approaches: 
\begin{itemize}
    \item {\it Case A:} we perform our analysis directly on the {\it observed} fiber-spectrum with no further corrections. This will provide galaxy properties corresponding to the inner parts of the galaxy, where ``inner'' depends on galaxy redshift and size, only integrating the fraction of the total galaxy light that is enclosed in the area covered by the fiber;
    \item{\it Case B:} we {\it rescale} the observed fiber-spectrum to the total photometry of the host galaxy before performing our analysis. This has the advantage over Case A that the total light of the galaxy is recovered,  but has the caveat of assuming that all stellar populations, mass to light ratios (M/L), emission line ratios, equivalent widths, etc. in the inner regions are similar than those at outer regions, ignoring radial variations (gradients);
    \item {\it Case C:} we perform our analysis on the {\it observed} fiber-spectra (not rescaled, Case A) and then we will apply {\it aperture corrections} to estimate global galaxy properties, which are in principle calibrated to account for differences in both brightness and stellar population parameters (age, metallicity, etc...).
\end{itemize}
We used our own code to extract the information from the SN host galaxy spectra and to derive the necessary parameters. These included the properties of the ionized gas and the underlying stellar populations present in the galaxy. The main procedures used for this analysis are described in detail in  \cite{2014A&A...572A..38G,2016A&A...591A..48G,2018ApJ...855..107G}, and summarised in sections \ref{sec:resc}-\ref{sec:aper}.

\begin{figure}[!t]
\includegraphics*[trim=0.75cm 0cm 1.75cm 1.25cm, clip=true,width=\columnwidth]{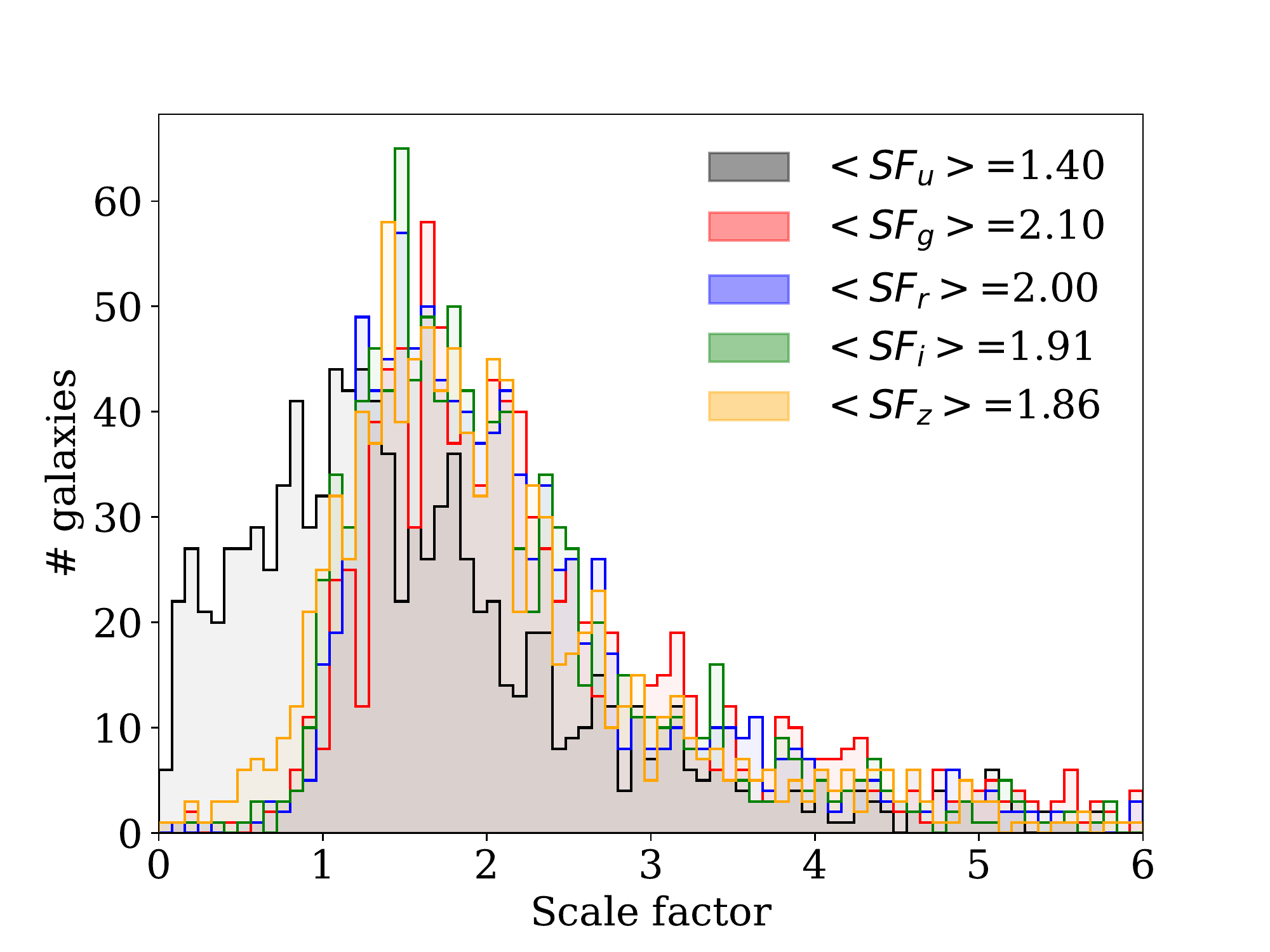}
\caption{Distributions of scale factors for the 1066 spectra of our initial sample. Each distribution represents a SDSS filter.}
\label{fig:sf}
\end{figure}

\subsection{Preprocessing: rescale to match SDSS photometry}\label{sec:resc}


Extensive galaxy parameters (scale dependent) can be corrected by scaling the synthetic broad band magnitudes measured from the spectrum  ({\it fibermag} in SDSS) to match the global photometric measurements ({\it modelmag} in SDSS), under the assumption that the mass to light ratio (M/L) and other properties such as colour, age, metallicity, and extinction obtained from the spectrum (hence representative of the area inside the fiber)  are the same as those outside the fiber.

\begin{figure*}
\includegraphics*[trim=0.4cm 0cm 0cm 0cm, clip=true,width=\textwidth]{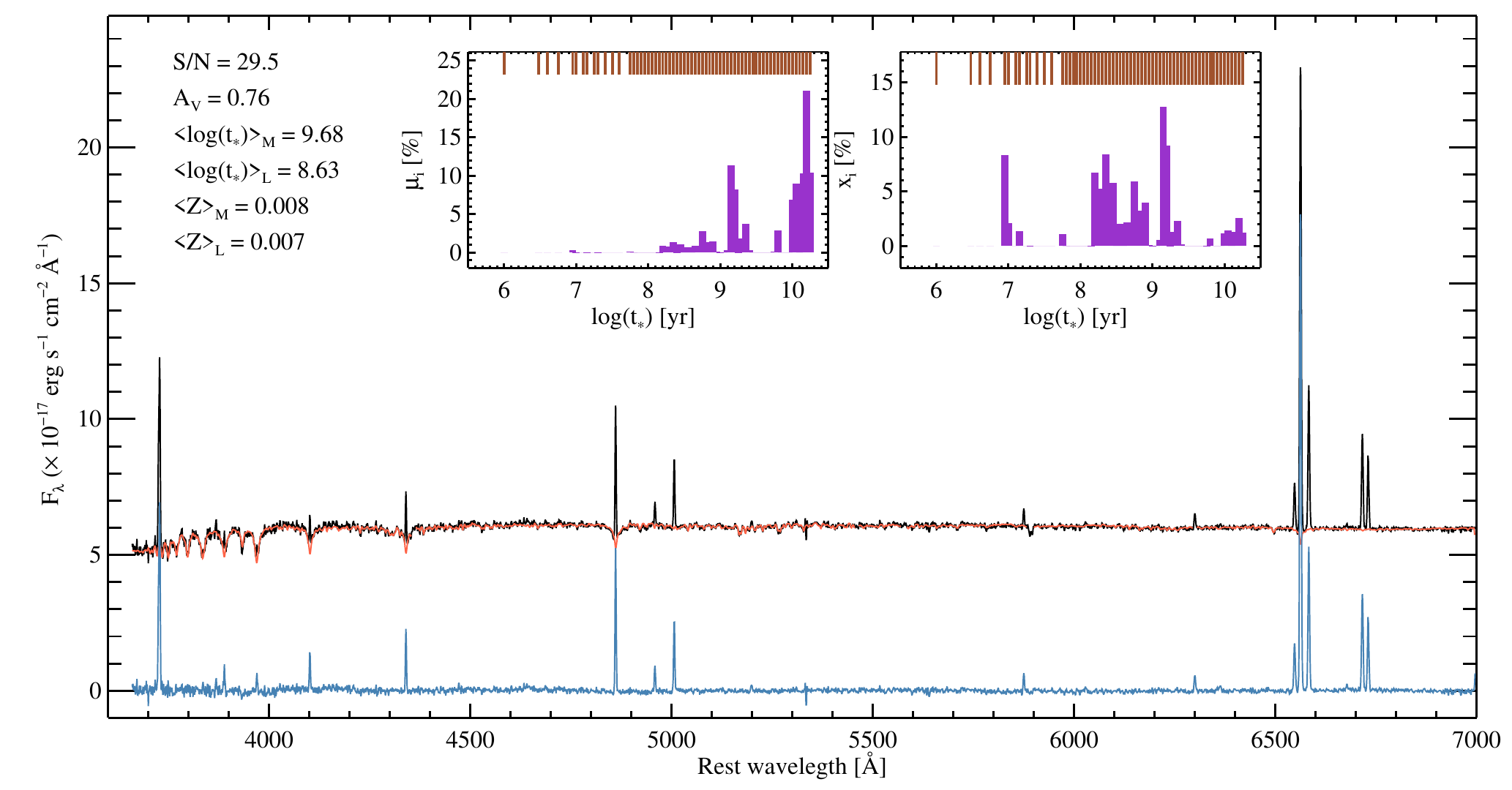}
\caption{Spectrum of SN2005ku (SDSS10805) host galaxy (black), together with the best {\tt STARLIGHT} fit (red) and the pure nebular emission line spectrum (the difference, in blue). Note that this method  accounts for the absorption produced at Balmer emission line positions. Inbox panels contain the distributions of the percentual contribution of the SSP models used in the STARLIGHT fit, weighted to the mass (left) and the light (right) of the host galaxy.}
\label{fig:spectrum}
\end{figure*}

We computed synthetic magnitudes by convolving the observed spectra with the response function of the $ugriz$ SDSS bands, and using the AB magnitude expression ($m$, \citealt{1983ApJ...266..713O,1996AJ....111.1748F}),
\begin{equation}
m=-2.5\log_{10}  \frac{\int{\lambda S(\lambda) f_\lambda(\lambda)d\lambda}}{c \int{\lambda^{-1}S(\lambda) d\lambda}} - zp,
\end{equation}
where $f_{\lambda}$ is the specific flux per unit wavelength, and the zero-point (zp = 48.6 mag) is chosen such that an object with a specific flux of 3631~Jy has $m=0$. The function $S(\lambda)$ already includes the transmissivity of the filter, the response of the instrument (telescope and detector), and the telluric features at some representative air-mass. $u$ and $z$ bands were in most cases outside the spectral coverage, so the scaling factors were computed by averaging the magnitude difference of only the three $gri$ bands and applying it to each spectrum. Figure \ref{fig:sf} shows the distribution of the computed scale factors in the 5 SDSS bands, where it can be seen the difference between the $u$ band and the other four.

We note that this rescaling does not affect other intensive quantities (scale independent), such as the stellar age and metallicity. Even by scaling the spectrum, the stellar populations that we infer from the spectrum are still representative of the area covered by the fiber, which might not be representative of the whole galaxy if radial gradients are present. 

As noted above, from here on we will apply the following steps described in the remaining of Section \ref{sec:HG} to both the observed (Case A) and to the rescaled spectra (Case B) independently. 

\subsection{Preprocessing: correct MW extinction and restframe}
 
All spectra (observed and rescaled) were then converted from vacuum to air wavelengths, and corrected for the Milky Way dust extinction using the dust maps of \cite{2011ApJ...737..103S} retrieved from the NASA/IPAC Infrared Science Archive (IRSA), applying the standard Galactic reddening law with \Rv = 3.1 \citep{1989ApJ...345..245C,1994ApJ...422..158O}. After that, they were also shifted to rest-frame wavelengths.

\subsection{Stellar population parameters}

Galaxy spectra can be divided in two components: the stellar continuum, and the ionized gas emission lines. Adopting the assumption that the star formation history (SFH) of a galaxy can be approximated as the sum of discrete star formation bursts, the observed stellar spectrum of a galaxy can be represented as the sum of spectra of single stellar population (SSP) with different ages and metallicities. To infer them, we used the {\tt STARLIGHT} code \citep{2009RMxAC..35..127C, 2005MNRAS.358..363C, 2006MNRAS.370..721M, 2007MNRAS.381..263A}. {\tt STARLIGHT} determines the fractional contribution of the different SSP models to the light, $x_i$ and to the galaxy mass, $\mu_i$. We can then estimate the mean light weighted (L) or mass weighted (M) age and metallicity of the stellar population from 
\begin{eqnarray}
\langle \log t_* \rangle_{L/M} & = & \sum_{i=1}^{N_*} w_i \log t_i, \\
\langle Z_* \rangle_{L/M}      & = & \sum_{i=1}^{N_*} w_i Z_i,
\end{eqnarray}
where $t_i$ and $Z_i$ are the age and the metallicity of the $i$-th SSP model, and  $w_i=x_i$ or $w_i=\mu_i$ for light- and mass-weighted quantities, respectively. Dust effects, parametrized by A$_V^*$, are modeled as a foreground screen with a \cite{1989ApJ...345..245C} reddening law assuming R$_V$ = 3.1.

We have used the $GM$ base of SSP models presented in \cite{2013A&A...557A..86C}. This base is a combination of SSP spectra from \cite{2010MNRAS.404.1639V} based on stars from the MILES library which start at an age of 63 Myr, with the models of \cite{2005MNRAS.357..945G} which relies on the synthetic stellar spectra from the Granada library for younger ages \citep{2005MNRAS.358...49M}. They are based on the \cite{1955ApJ...121..161S} Initial Mass Function (IMF) and the evolutionary tracks of \cite{2000A&AS..141..371G}, except for the youngest ages ($<$3 Myr), which are based on Geneva tracks \citep{1992A&AS...96..269S, 1993A&AS..102..339S, 1993A&AS...98..523S, 1993A&AS..101..415C}. We selected a set of 248 SSP models with 62 different ages (from 1~Myr to 14~Gyr) for each of the four metallicities (0.2, 0.4, 1.0 and 1.5 solar).

\cite{2013A&A...557A..86C} compared the stellar parameters obtained using different SSP bases, and the overall conclusion is that, from a statistical perspective, STARLIGHT results do not depend strongly on the choice of SSP models. Mean ages and extinctions all agree to within relatively small margins, and the same is true for stellar masses (once built-in differences in IMFs are accounted for).

Some wavelength regions containing known optical nebular emission lines, telluric absorptions, or strong night-sky emission lines were masked out from the fit. As an example, Figure~\ref{fig:spectrum} shows the observed central spectrum of the host galaxy of SN 2005ku and the {\tt STARLIGHT} fit. The fit subtraction gives the pure emission line spectrum that is used to measure the ionized gas emission lines. 

In this work we focus on studying those parameters for which we have aperture corrections, so we have only made use of the stellar mass ($M_*$) from the STARLIGHT output. Mass uncertainties were determined from repeating the fitting on 1000 realisations of the integrated spectrum, sampled from its flux and error, by adopting the width of the distribution of these 1000 values. We note that for the spectra scaled to the photometry, only extensive (scale dependent) quantities change from the values found from the fits to the observed spectra. Intensive quantities are equivalent (See Appendix \ref{sec:app} for a comparison of the stellar parameters of Case A and B).

\begin{figure}
\includegraphics*[trim=0.0cm 0cm 0.0cm 0cm, clip=true,width=\columnwidth]{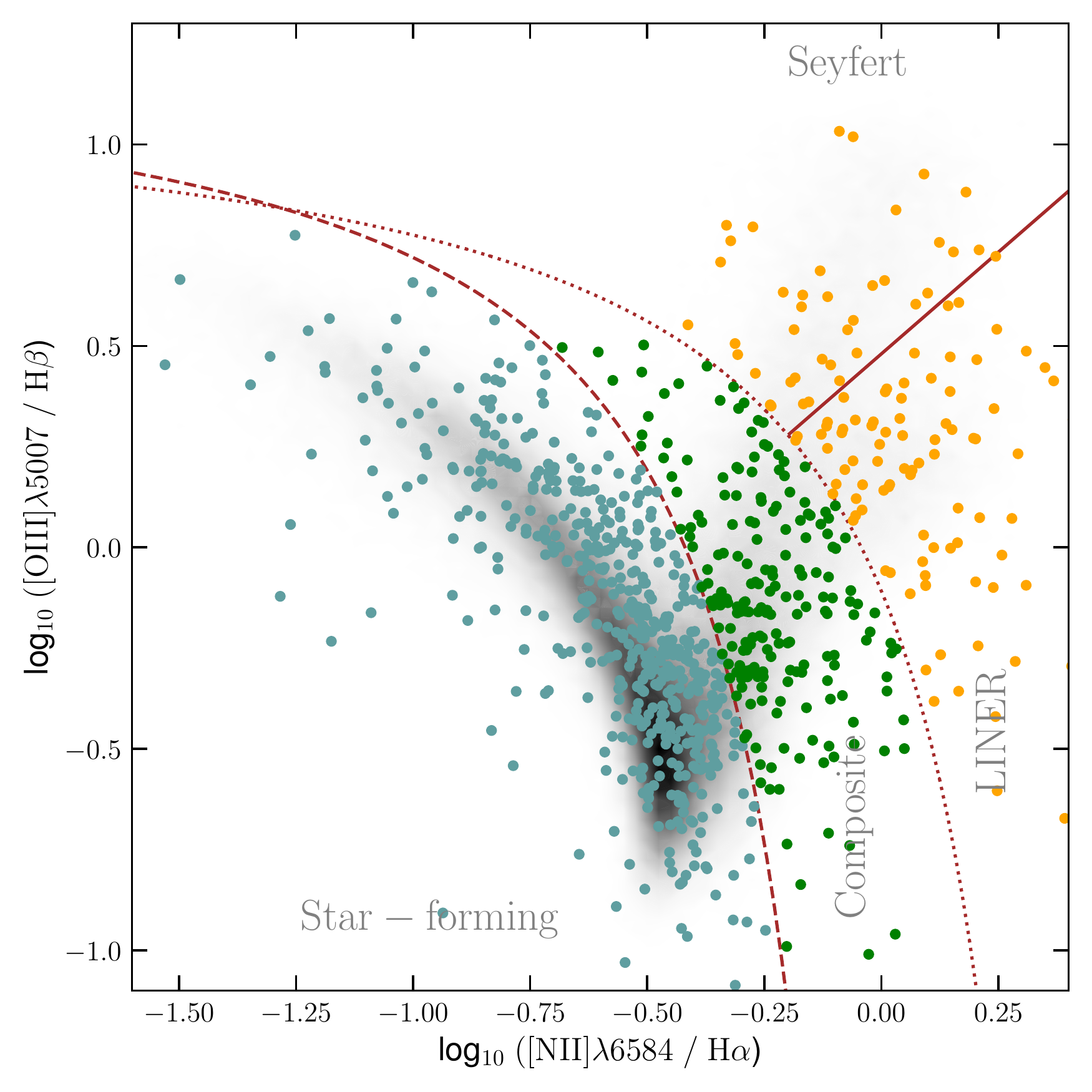}
\caption{[\ion{O}{iii}]/H$\beta$ vs [\ion{N}{ii}]/H$\alpha$ BPT diagram of the 869 galaxies with reliable emission lines. We excluded all 137 galaxies (in orange) that fall in the region above the line by \cite{2001ApJ...556..121K}, which determines that the ionization source is not underlying star-formation but active galactic nuclei.}
\label{fig:bpt}
\end{figure}

\subsection{Ionized gas parameters} \label{sec:el}

Subtracting the STARLIGHT fits from the observed spectra, we obtain the pure nebular emission line spectra, and accurately measure the flux of the most prominent emission lines (H$\beta$, [\ion{O}{iii}] $\lambda$5007, H$\alpha$, [\ion{N}{ii}] $\lambda$6583) by fitting a weighted nonlinear least-squares fit with a single Gaussian plus a linear term. The uncertainty of the flux was determined from the S/N of the measured line flux and the ratio between the fitted amplitude of the Gaussian to the standard deviation of the adjacent continuum. Monte Carlo simulations were performed to obtain realistic errors to the line fluxes from these two measurements. For full details of the procedure see Appendix C in \cite{2012A&A...545A..58S}.

We need to apply some quality cuts to our measurements. From the initial 1066 spectra, we only kept those where we measured a S/N > 2 simultaneously for all four emission lines listed above (used in the oxygen abundance estimation). In addition, we also kept galaxies whose spectra had an H$\alpha$ and H$\beta$ with a S/N > 2  independently of the S/N of the other two emission lines (used in the $A_V$ and SFR measurements). In this way, we did not miss galaxies with strong H$\alpha$ that can in turn be corrected for extinction using the Balmer decrement reliably. We note that a galaxy with strong H$\alpha$ and faint [\ion{N}{ii}]$\lambda$6583 is not suspicious of being ionized by AGN emission (very low [\ion{N}{ii}]/H$\alpha$; see Section \ref{sec:agn}). These combined criteria allows us to be secure not to miss any bright H$\alpha$ line to measure the SFR, and at the same time not miss objects with reliable oxygen abundance. We visually confirmed that these cuts leave out spectra where lines are not reliably detected. There are 869 spectra passing the above cuts. For these objects the H$\alpha$ equivalent width (H$\alpha$EW) was measured from the normalized spectra, resulting from the division of the actual spectra by the STARLIGHT fit. While the H$\alpha$ flux is an indicator of the ongoing SFR traced by ionizing OB stars, the H$\alpha$EW measures how strong this is compared to the stellar continuum, which is dominated by old low-mass non-ionizing stars and therefore accounts for most of the galaxy stellar mass. H$\alpha$EW can be thought of as an indicator of the strength of the ongoing SFR compared with the past SFR, which decreases with time if no new stars are created, and it is thus a reliable proxy for the age of the youngest stellar components \citep{2009A&A...508..615L,2016A&A...593A..78K}.

The observed ratio of \Had\ and \Hbd\ emission lines provides an estimate of the gas-phase dust attenuation $A_V^g$ along the line of sight through a galaxy. Assuming an intrinsic ratio  $I$(\Had)/$I$(\Hbd)=2.86, valid for case B recombination with $T=10,000$~K and electron density 10$^2$~cm$^{-3}$ \citep{2006agna.book.....O}, and using the \cite{1989ApJ...345..245C} extinction law with the updated \cite{1994ApJ...422..158O} coefficients, we can estimate \EBV. With this and adopting the Milky Way average value of \Rv = $A_V^g$ / \EBV = 3.1, we calculated $A_V^g$. The emission lines previously measured were corrected for the dust extinction before calculating any further measurement. 

\subsubsection{Removing AGN contribution}\label{sec:agn}

Some methods used to derive relevant quantities can only be applied if the ionization source exclusively arises from the stellar radiation. To identify AGN contamination in the galaxy centers we used the BPT diagnostic diagram \citep{1981PASP...93....5B, 1987ApJS...63..295V}, a map of $O3\equiv\log_{10}\left(\frac{\textrm{\OIIId}}{\textrm{\Hbd}}\right)$, and $N2\equiv\log_{10}\left(\frac{\textrm{\NIId}}{\textrm{\Had}}\right)$, on which gas ionized by different sources occupies different areas. Two criteria commonly used to separate star-forming (SF) from AGN-dominated galaxies are the expressions in \cite{2001ApJ...556..121K} and \cite{2003MNRAS.346.1055K}. However, it should be noted that the latter is an empirical expression,  and {\it bona fide} {\sc \ion{H}{ii}} regions can be found in the composite area determined by it \citep{2014A&A...563A..49S}. Galaxies with emission line ratios falling in the AGN-dominated region according to the criterion of \cite{2001ApJ...556..121K} were excluded from the following analysis. This reduced the sample by 137 objects to 732 galaxies (see Figure \ref{fig:bpt}).

\subsubsection{Star formation rate}\label{sec:sfr_det}

We estimated the ongoing star formation rate (SFR) from the extinction-corrected H$\alpha$ flux $F(\mathrm{H}\alpha)$ using the expression given by \citet{1998ApJ...498..541K}:
\begin{equation}
\mathrm{SFR}\,[M_{\sun}\,\mathrm{yr}^{-1}]=7.9365\times 10^{-42}\,L(\mathrm{H}\alpha),
\end{equation}
where $L(\mathrm{H}\alpha)$ is the H$\alpha$ luminosity in units of erg~s$^{-1}$. \cite{2015A&A...584A..87C} demonstrated that the H$\alpha$ luminosity alone can be used as a tracer of the current SFR, even without including UV and IR measurements, once the underlying stellar absorption and the dust attenuation effects have been accounted for, as we did here. This measurement is also used to estimate the specific SFR (sSFR = SFR / Mass). 

\subsubsection{Oxygen abundance}\label{sec:oh_det}

\begin{table*}[!t]
\caption{Results of the fifth-order polynomial fits to the fixed angular aperture corrections given in \cite{2013A&A...553L...7I} and \cite{2016ApJ...826...71I}, in the form $a_0 + a_1*x + a_2*x^2 + a_3*x^3 + a_4*x^4 + a_5*x^5$, where $x$ is $r/R_{50}$.}
\begin{center}
\begin{tabular}{lcccccc}
\hline
Parameter         &  $a_0$  &  $a_1$  &  $a_2$  &  $a_3$   &  $a_4$   &  $a_5$    \\
\hline
H$\alpha$         &  0.0000 &  0.1599 &  0.5027 & -0.2276  &  0.0167  &  0.0037   \\
H$\alpha$/H$\beta$&  1.1253 &  0.0274 & -0.3125 &  0.2959  & -0.1091  &  0.0143   \\
N2                &  0.0812 &  0.0477 & -0.2779 &  0.2715  & -0.1075  &  0.0153   \\
O3N2              & -0.1299 & -0.0366 &  0.2645 & -0.2624  &  0.1092  & -0.0164   \\ 
H$\alpha$EW       &  0.6262 &  0.4492 & -0.1635 &  0.03718 & -0.01926 &  0.004537 \\
\hline
\end{tabular}
\end{center}
\label{tab:apertures}
\end{table*}

Since oxygen is the most abundant metal in the gas phase and exhibits very strong nebular lines in optical wavelength, it is usually chosen as a metallicity indicator in ISM studies.

The most accurate method to measure ISM abundances (the so-called direct method) involves determining the ionized gas electron temperature, T$_e$, which is usually estimated from the flux ratios of auroral to nebular emission lines, e.g. \mbox{\OIIIt/\OIIIu}~(\citealt{2006A&A...448..955I, 2006A&A...454L.127S}). However, the temperature-sensitive lines such as \OIIIt~are very weak and difficult to measure, especially in metal-rich environments that correspond to lower temperatures. Since in our data the direct method cannot be used everywhere reliably (only 16 detections with S/N $>$ 5), we used instead other strong emission line methods to determine the gas oxygen abundance.

Many such methods have been developed throughout the years. The theoretical methods are calibrated by matching the observed line fluxes with those predicted by theoretical photoionization models. The empirical methods, on the other hand, are calibrated against \ion{H}{ii} regions and galaxies whose metallicities have been previously determined by the direct method. Unfortunately, there are large systematic differences between methods, which translate into a considerable uncertainty in the absolute metallicity scale (\citealt{2010A&A...517A..85L} and \citealt{2012MNRAS.426.2630L} for a review), while relative metallicities generally agree. The cause of these discrepancies is still not well-understood, although the empirical methods may underestimate the metallicity by a few tenths of dex, while the theoretical methods overestimate it \citep{2007RMxAC..29...72P, 2010ApJS..190..233M,2014MNRAS.441.2663P}.

We used the O3N2 empirical method to compute the elemental abundances in all galaxies. It was firstly introduced by \cite{1979A&A....78..200A}, as the difference between the two line ratios used in the BPT diagram (\OIIIHb~and \NIIHa). We here take the calibration introduced by \citealt{2004MNRAS.348L..59P} (hereafter PP04),
\begin{equation}
12 + \log~({\rm O/H}) = 8.73 - 0.32 \times{\rm O3N2}.
\end{equation}
This method has the advantage (over other methods) of being insensitive to extinction due to the small separation in wavelength of the emission lines used for the ratio diagnostics thus minimising differential atmospheric refraction (DAR). It has been found to be valid for O3N2$<$2 (12+$\log_{10}$O/H$>$8.09). The uncertainties in the measured metallicities were computed by including the statistical uncertainties of the line flux measurements and those in the derived galaxy reddening, and by propagating them into the metallicity determination. 

In summary, for all 732 remaining galaxies we derived: A$_V^g$, H$\alpha$EW, SFR, sSFR, and O/H.

\begin{figure*}[!t]
\centering
\includegraphics*[trim=0.0cm 0.0cm 0.0cm 0.0cm, clip=true,width=0.89\textwidth]{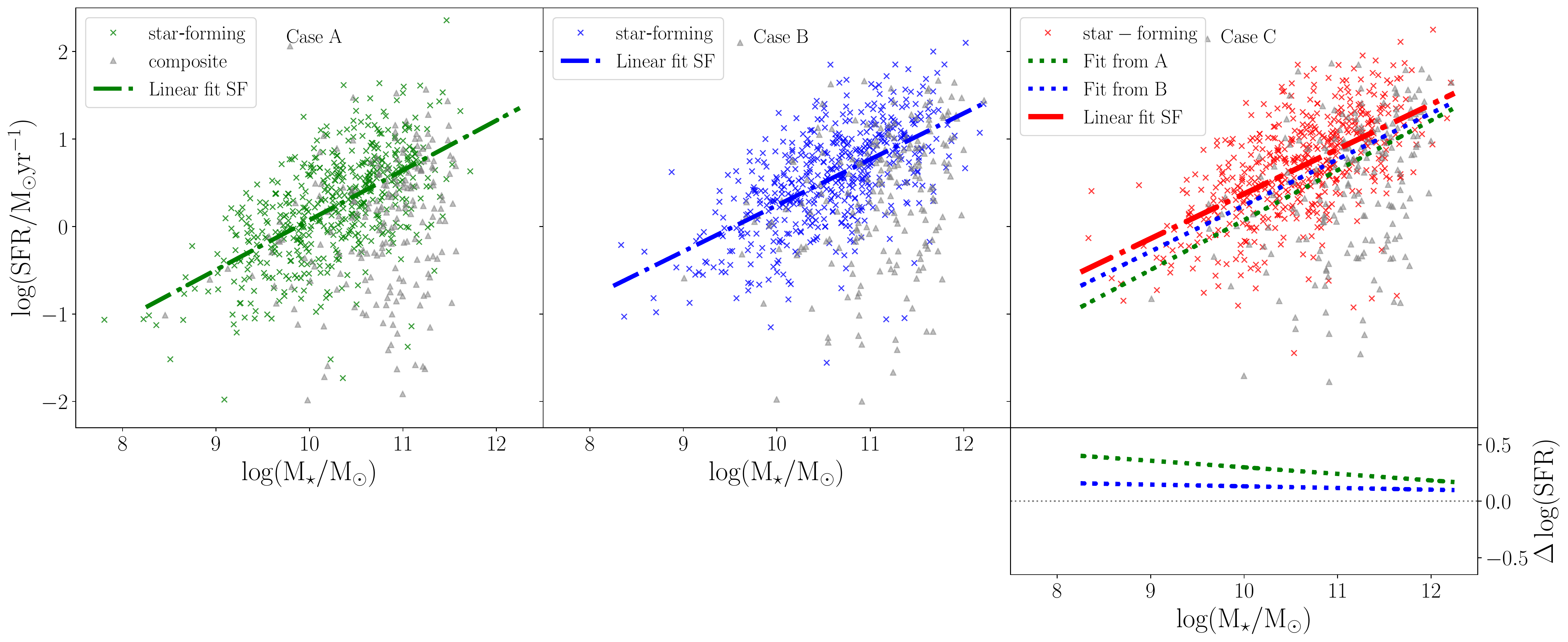}
\includegraphics*[trim=0.0cm 0.0cm 0.0cm 0.0cm, clip=true,width=0.89\textwidth]{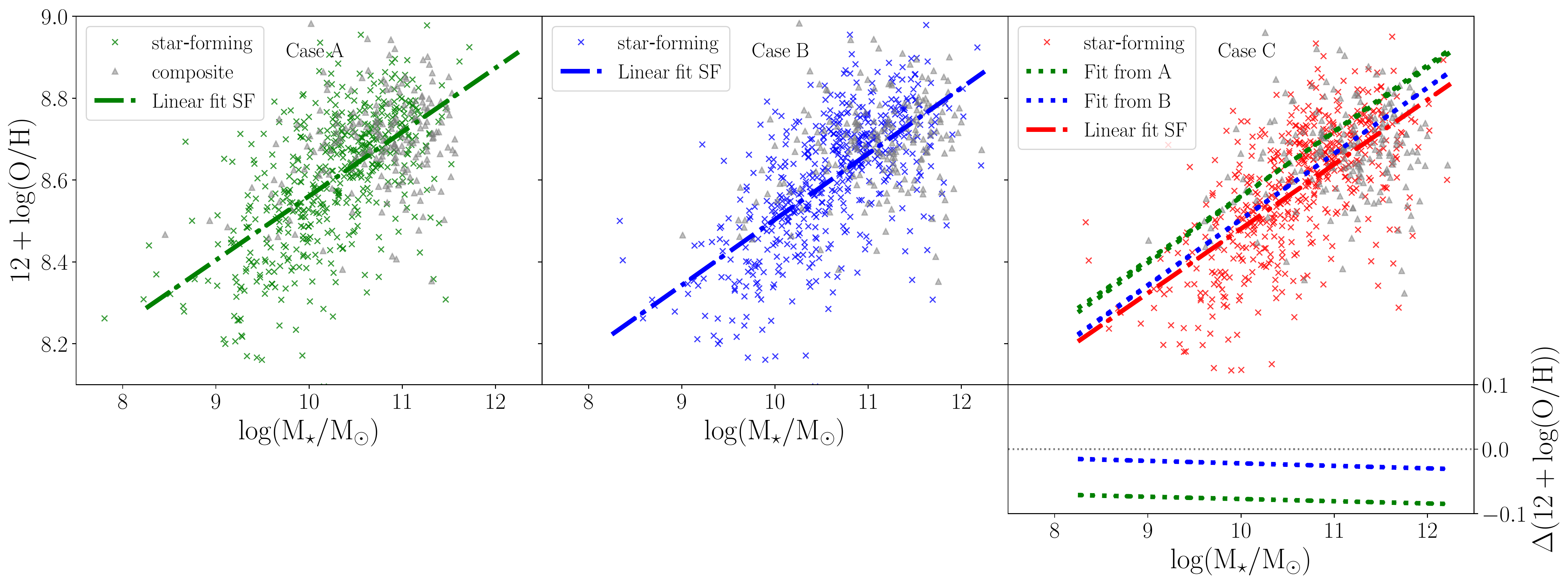}
\includegraphics*[trim=0.0cm 0.0cm 0.0cm 0.0cm, clip=true,width=0.89\textwidth]{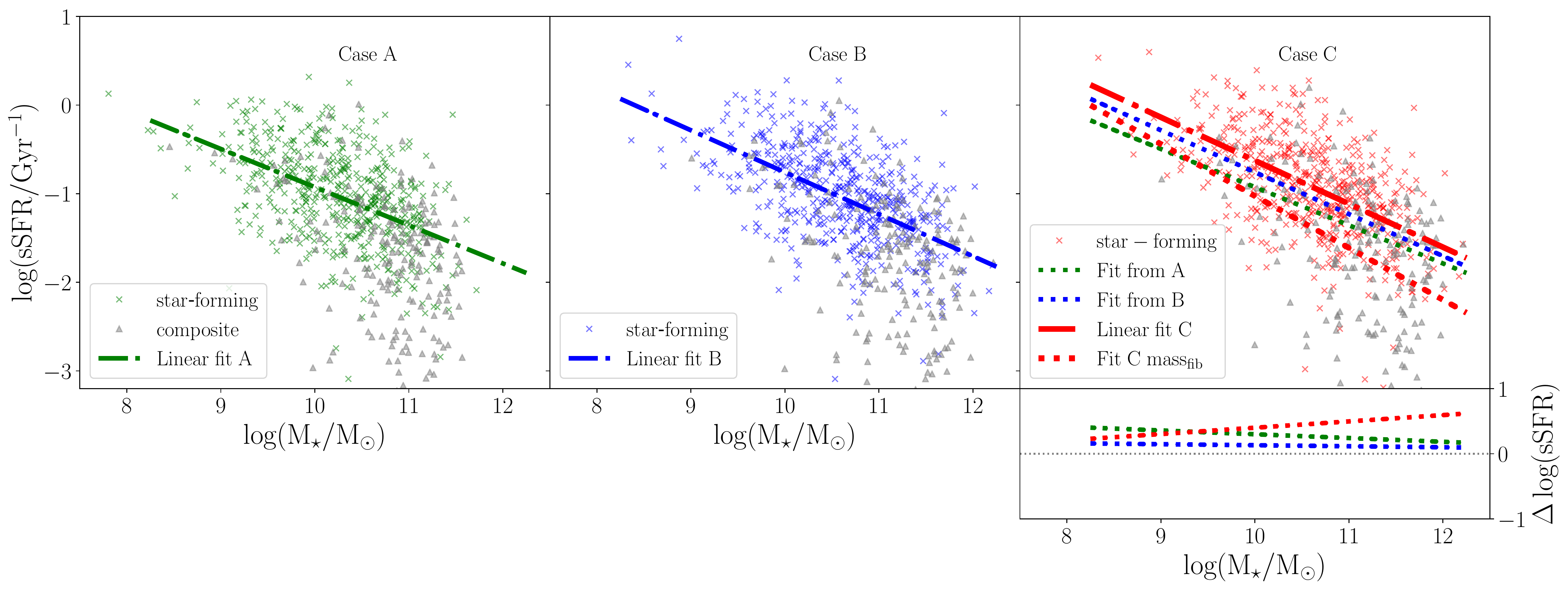}
\caption{SFR--M$_\star$, O/H--M$_\star$, and sSFR--M$_\star$ relations for our sample of galaxies. Star-forming (colored crosses) and composite (grey triangles)  based on the BPT classification are represented in each diagram. Left and central panels correspond to Case A and Case B, respectively. We included the linear fit of the SFR--M$_\star$ relation for all the star-forming galaxies. Right panels correspond to Case C, the aperture corrected. In the upper panel we included the linear fit of the SFR--M$_\star$ relation for all the star-forming galaxies considered in case C (dashed red line), the fit from Case A (dotted green line), and the fit from Case B (dotted blue line). In the bottom-right panel, sSFR--M$_\star$ relation, we also included a fourth relation (dotted red) corresponding to an aperture corrected SFR (Case C) normalized to the mass measured in the fiber spectrum (case A). In the lower panels we show the difference between the linear fit in case C and the linear fits of Case A (dotted green line) and Case B (dotted blue line).}
\label{fig:mass-relations}
\end{figure*}

\begin{figure*}[!t]
\includegraphics*[trim=0.0cm 0.0cm 0.0cm 0.0cm, clip=true,width=0.48\textwidth]{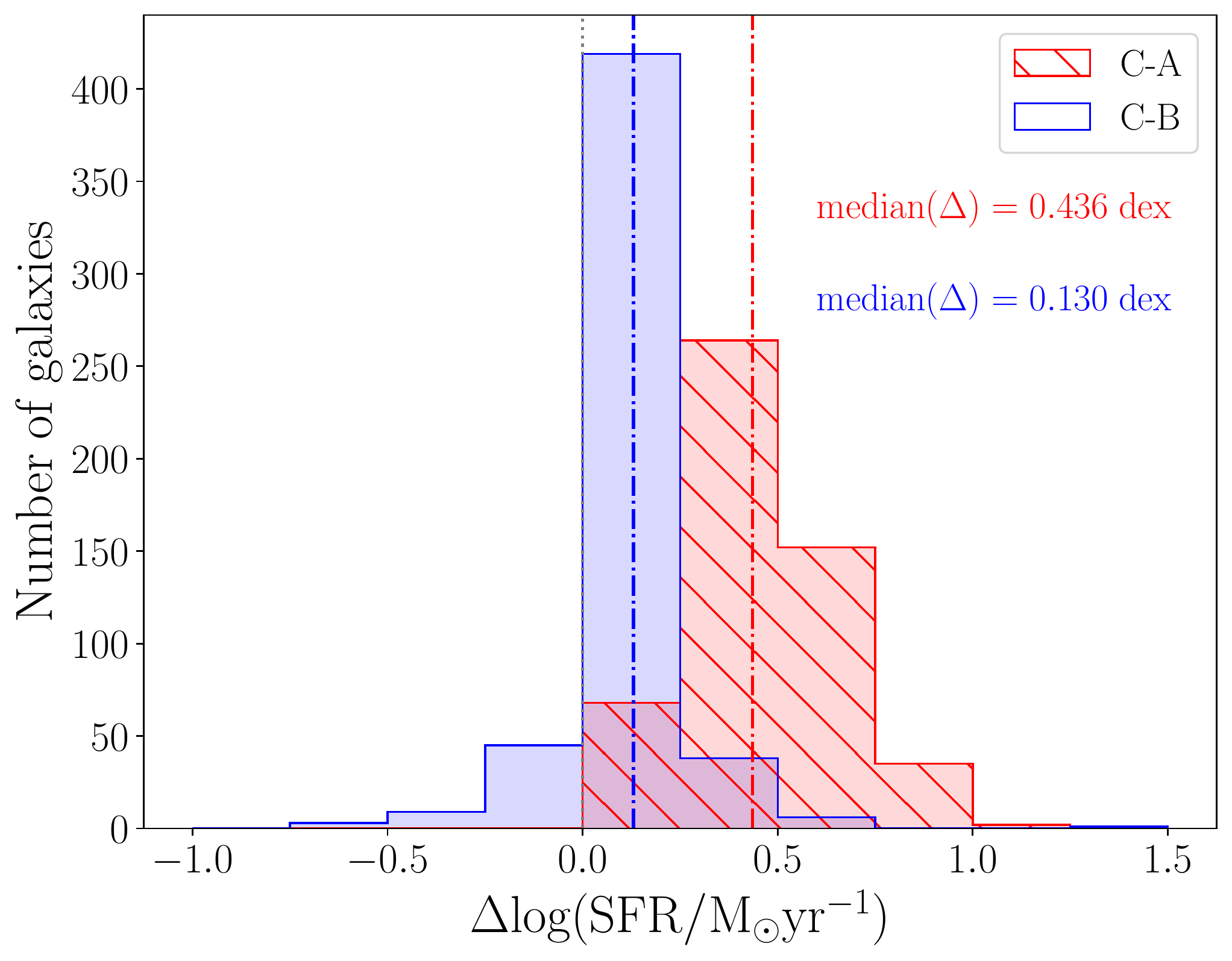}
\includegraphics*[trim=0.0cm 0.0cm 0.0cm 0.0cm, clip=true,width=0.48\textwidth]{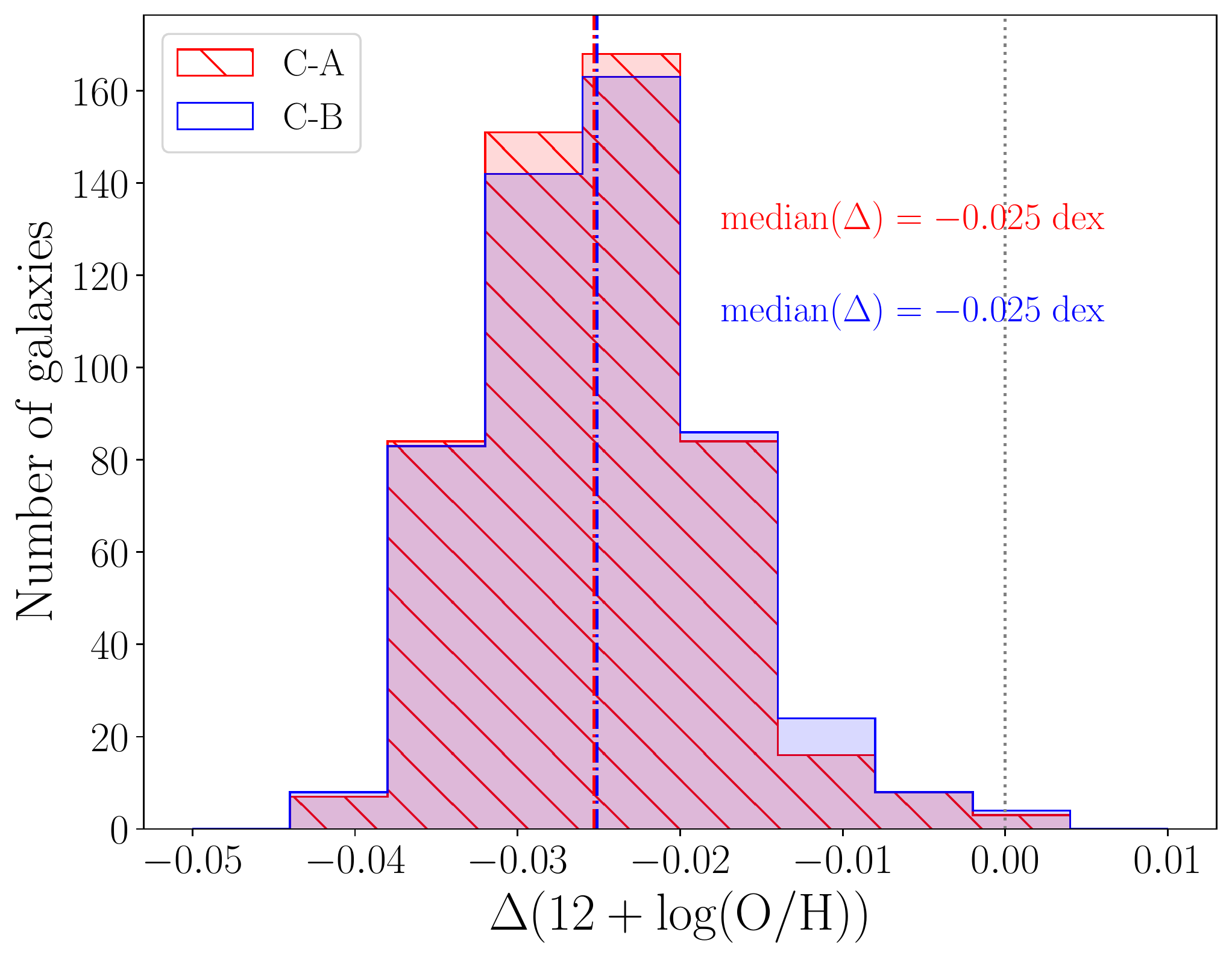}\\
\includegraphics*[trim=0.0cm 0.0cm 0.0cm 0.0cm, clip=true,width=0.48\textwidth]{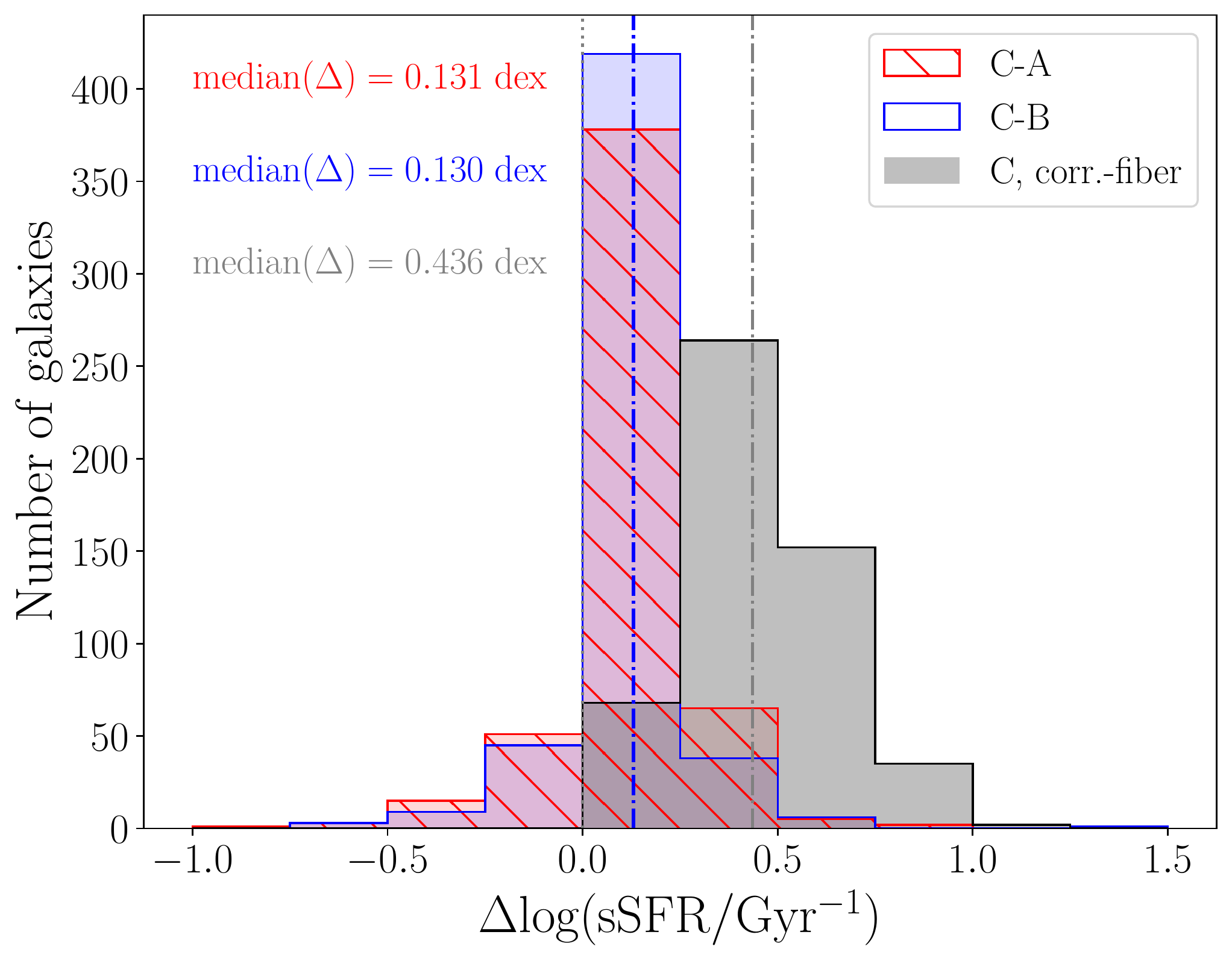}
\includegraphics*[trim=0.0cm 0.0cm 0.0cm 0.0cm, clip=true,width=0.48\textwidth]{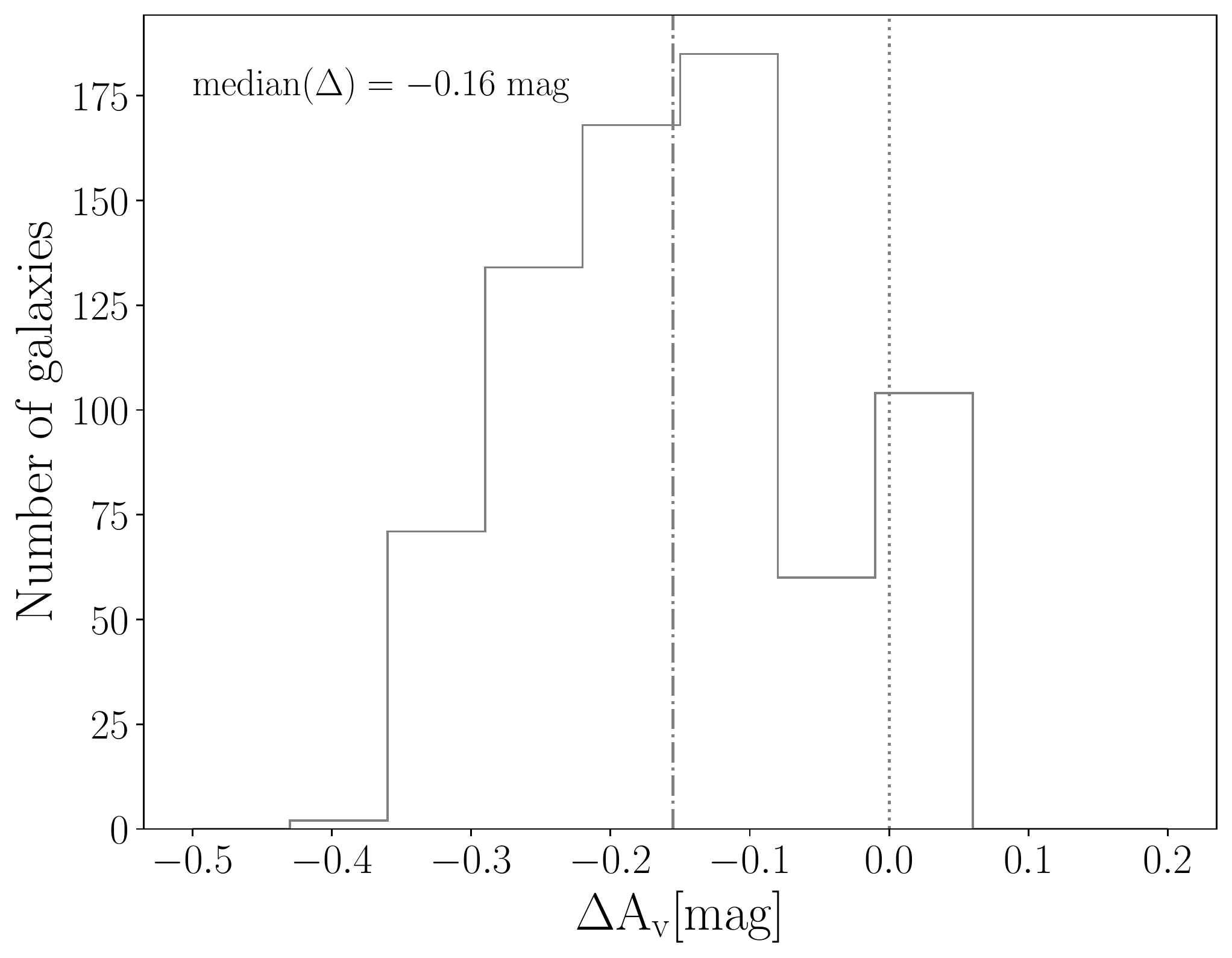}
\caption{Upper left panel: histogram of the differences in SFR values for cases C and A (red) and cases C and B (blue); upper right panel: histogram of the differences in 12+log(O/H) values for cases C and A (red) and cases C and B (blue); lower left panel: histogram of the differences in sSFR values for cases C and A (red), cases C and B (blue), and with the SFR corrected but divided by the observed mass within the fiber (grey); and lower right panel: histogram of the differences in A$_v$ corrected and uncorrected for aperture. The vertical dashed lines represent the median value of this difference (related to the color of each histogram) and the vertical dotted lines show zero value on the x-axis.}
\label{fig:hist-diff-sfr}
\end{figure*}


\begin{figure*}[!t]
\centering
\includegraphics*[trim=0.7cm 0cm 0.55cm 0cm, clip=true,width=0.33\textwidth]{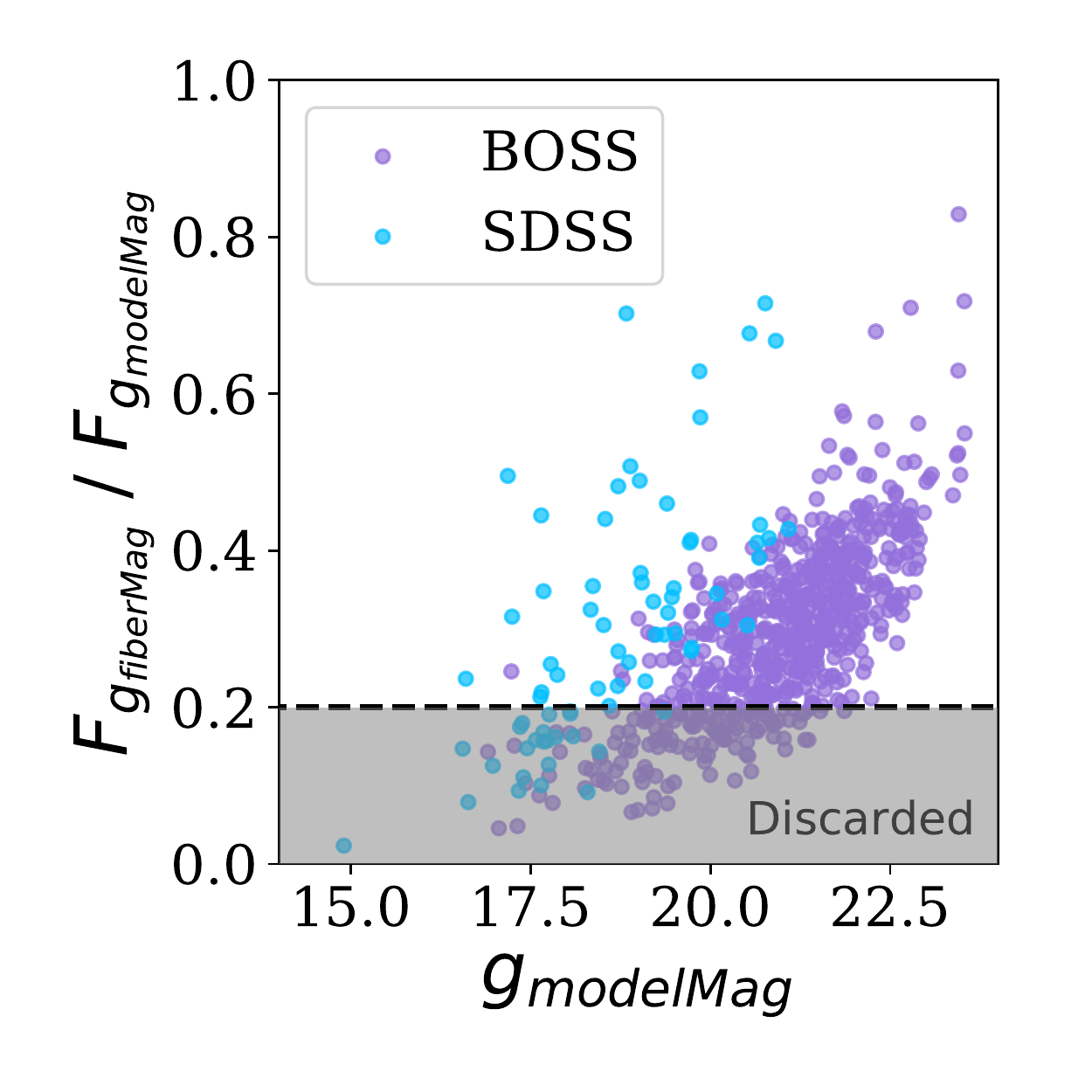}
\includegraphics*[trim=0.7cm 0cm 0.55cm 0cm, clip=true,width=0.33\textwidth]{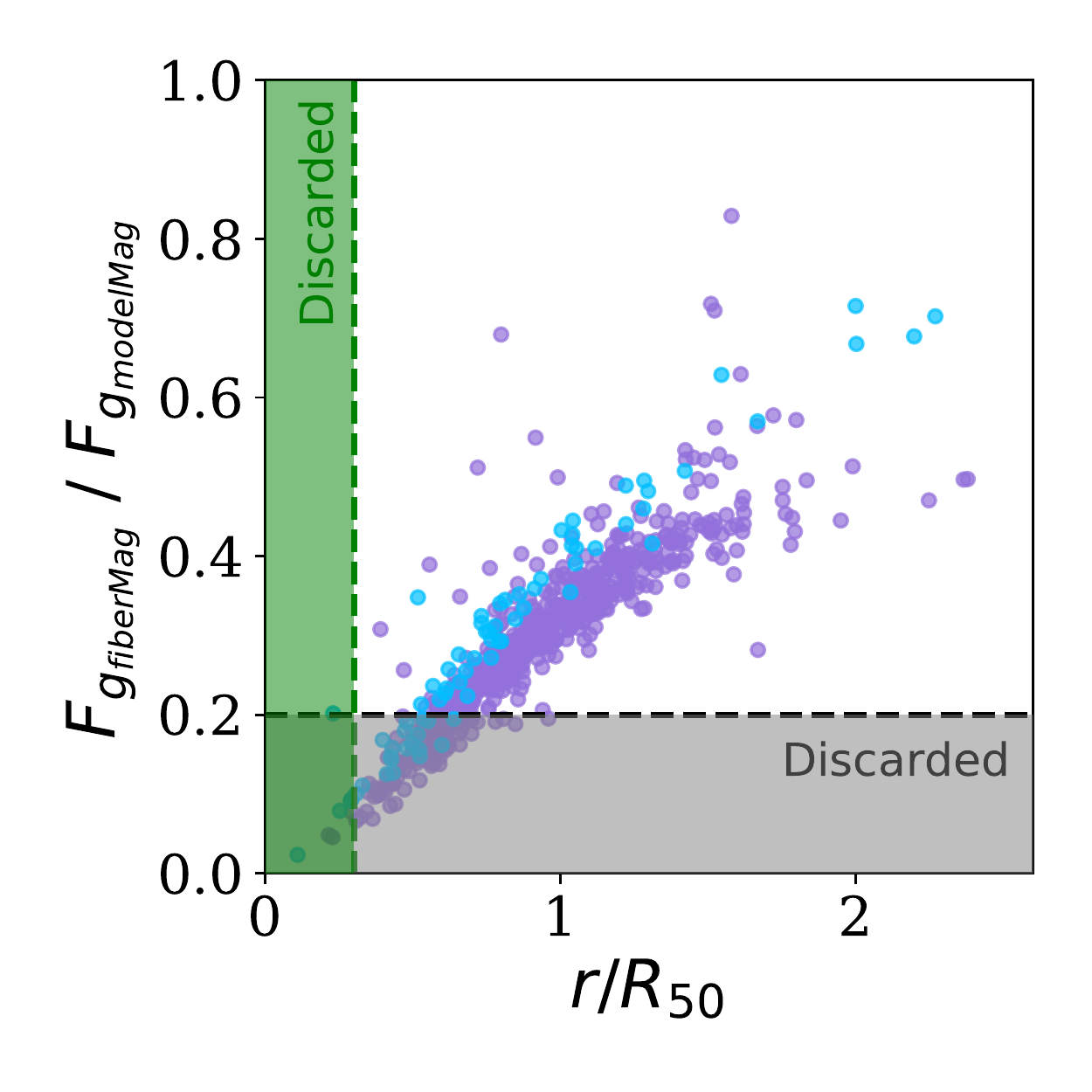}
\includegraphics*[trim=0.7cm 0cm 0.55cm 0cm, clip=true,width=0.33\textwidth]{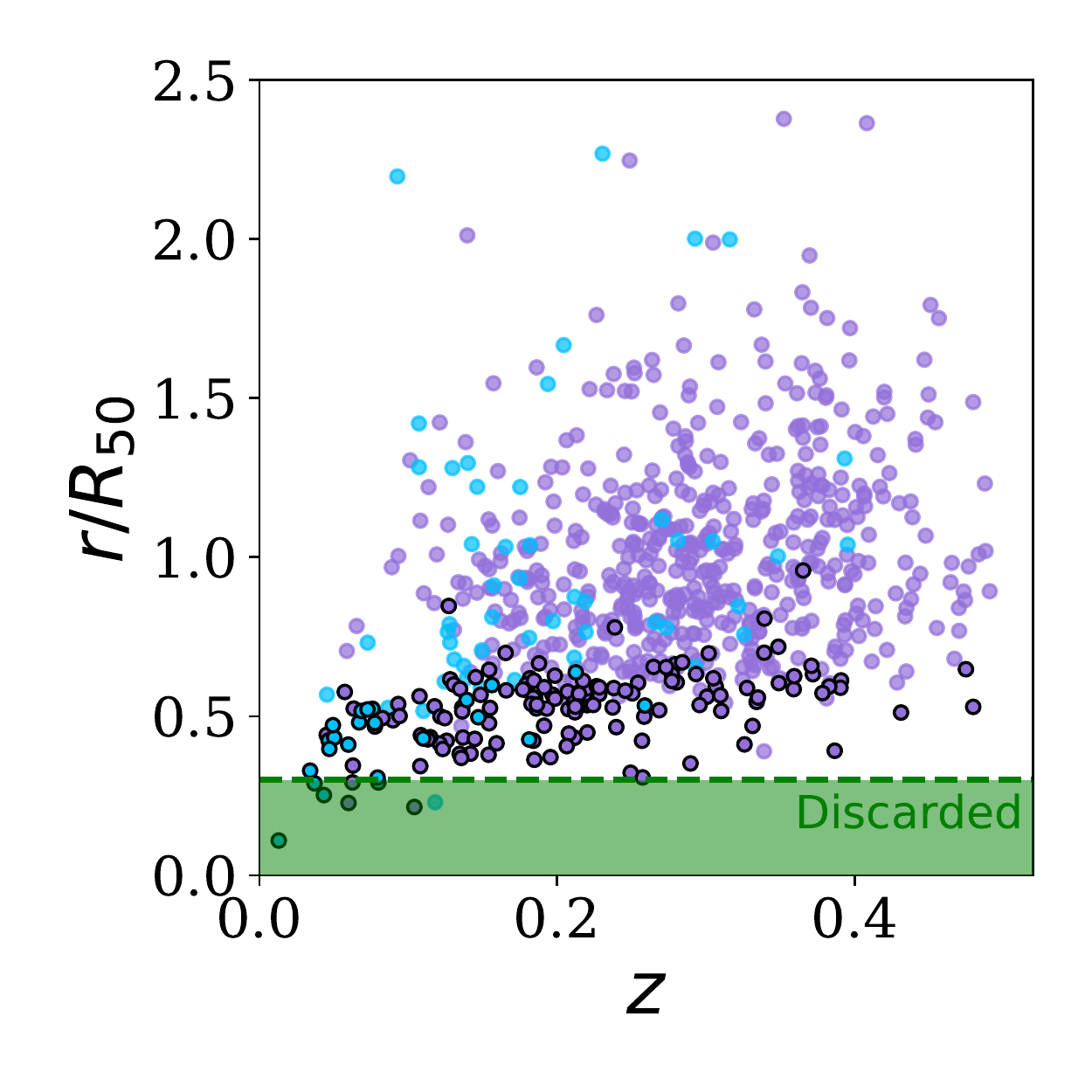}
\caption{
{\it Left:} $g$-band flux fraction covered within the fiber with respect to the total magnitude ($modelMag$ in the SDSS catalogue) of the galaxy. The horizontal line at 0.2 corresponds the cut suggested by \cite{2005PASP..117..227K} and used in \cite{2011ApJ...743..172D} and \cite{2016ApJ...821..115W} to justify that objects below the line on the shaded region have spectra that is not representative of the whole galaxy, and therefore were excluded in those analyses. {\it Middle:} $g$-band fiber flux fraction vs. radii of the galaxy normalized to the Petrosian R$_{50}$ radius (radius of a circle that contains 50\% of the galaxy light). The horizontal line is the \cite{2005PASP..117..227K} limit, where all objects below the line in the shaded region would be excluded. The vertical line corresponds to the range where our aperture corrections are applied, objects on the left shaded region (r/R$_{50}$ $<$ 0.3) are excluded because \cite{2013A&A...553L...7I,2016ApJ...826...71I} corrections cannot be applied. We note that those galaxies with r/R$_{50}$ $\ge$ 0.3 (right from the vertical line) but g-band fiber flux fraction lower than 0.2 (below the horizontal line) would have been excluded but are included in this work. {\it Right:} Normalized radii (r/R$_{50}$) of the galaxies vs. redshift. Dots with black edges correspond to the 159 objects that we would have discarded using the $g$-band fraction criterion.
In our selection we have mostly lost galaxies at low redshifts.}
\label{fig:apercor2}
\end{figure*}

\subsection{Aperture corrections}\label{sec:aper}

Galaxy spectra taken with fiber or slit spectroscopy are not always comparable because different fractions are sampled, and the fraction depends on the size and distance (redshift) of each galaxy. This aperture effect is most noticeable in the low-redshift Universe, affecting the measurements of some galaxy properties, especially extensive properties (e.g. SFR) that can only increase as more light is integrated. On the other hand, the variation expected for intensive properties (e.g. O/H, A$_V$) is smaller. 

One would be able to correct for the missing light by studying a sample of galaxies for which different fractions can be extracted and the desired parameters measured. Then, one can study how these parameters change as the integrated extent of the galaxy varies. IFS is the perfect technique to approach these problems, since one can obtain multiple spectra mapping the whole extent of a galaxy, and simulate several spectral extractions. \cite{2013A&A...553L...7I} estimated empirical aperture corrections from nearby galaxies observed by the CALIFA Survey \citep{2012A&A...538A...8S}. Spectra of increasing apertures were extracted from a representative sample of 165 CALIFA galaxies, and the growth of both H$\alpha$, H$\alpha$EW and the ratio of H$\alpha$/H$\beta$ were studied for different galaxy morphologies, inclinations and masses. Further analysis with a larger sample on the growth of the ratio of [\ion{N}{ii}]/H$\alpha$ and [\ion{O}{iii}]/H$\beta$, the line ratios used for the estimation of the oxygen abundance with the O3N2 calibrator, was also presented in \cite{2016ApJ...826...71I} and \cite{2017A&A...599A..71D}. These works provided growth curves for all these measurements normalized to the total integrated value at 2.5 R$_{50}$, where R$_{50}$ is the Petrosian radius containing 50\% of the total galaxy flux in a particular band. The growth curves used in this work described as fifth-order polynomials are summarized in Table \ref{tab:apertures}.

These aperture corrections have two caveats: (i) for large and very nearby galaxies the fiber cover only a very small fraction r/R$_{50}$ of the galaxy and aperture corrections would have huge errors. For this reason Iglesias-P\'aramo et al. give tabulated values starting from r/R$_{50}$ = 0.3; (ii) on the other hand, for apparently small and/or high redshift galaxies for which the galaxy size is comparable to the seeing of the observation, R$_{50}$ is not well defined since it is measuring the half light of the PSF and cannot account for the shape of the galaxy. For such a small/far galaxies, no aperture correction is needed. 

The final goal of applying these aperture corrections is to be able to make proper comparisons among galaxies for all parameters within a range of redshifts, as if the fiber covered the same extent in all galaxies. Here we made use of the information presented in these two studies to correct the host galaxy parameters measured in previous sections. For that, we obtained the R$_{50}$ parameter in the $r$-band from the SDSS database, and calculated the area covered by either the SDSS or the BOSS fiber  as a fraction of R$_{50}$. Therefore, by comparing the coverage for each galaxy to the tabulated values, one can apply a simple correction factor to the values estimated in the previous sections.

Considering only those galaxies which fulfilled the condition of having a coverage of r $>$ 0.3 R$_{50}$, and with the values for the fiber radius being 1" for BOSS and 1.5" for SDSS, only 8 galaxies were removed, so we kept 724 galaxies.

In Appendix \ref{app:tab} we  provide the final measurements of the main properties for the 3 cases (A, B, C) that will be used in the following sections. These include the scaled galaxy stellar mass, the SFR, the oxygen abundance, the H$\alpha$EW and the A$_V^g$.

\begin{figure}[!t]
\includegraphics[trim=0cm 0cm 0cm 0cm, clip=true,width=\columnwidth]{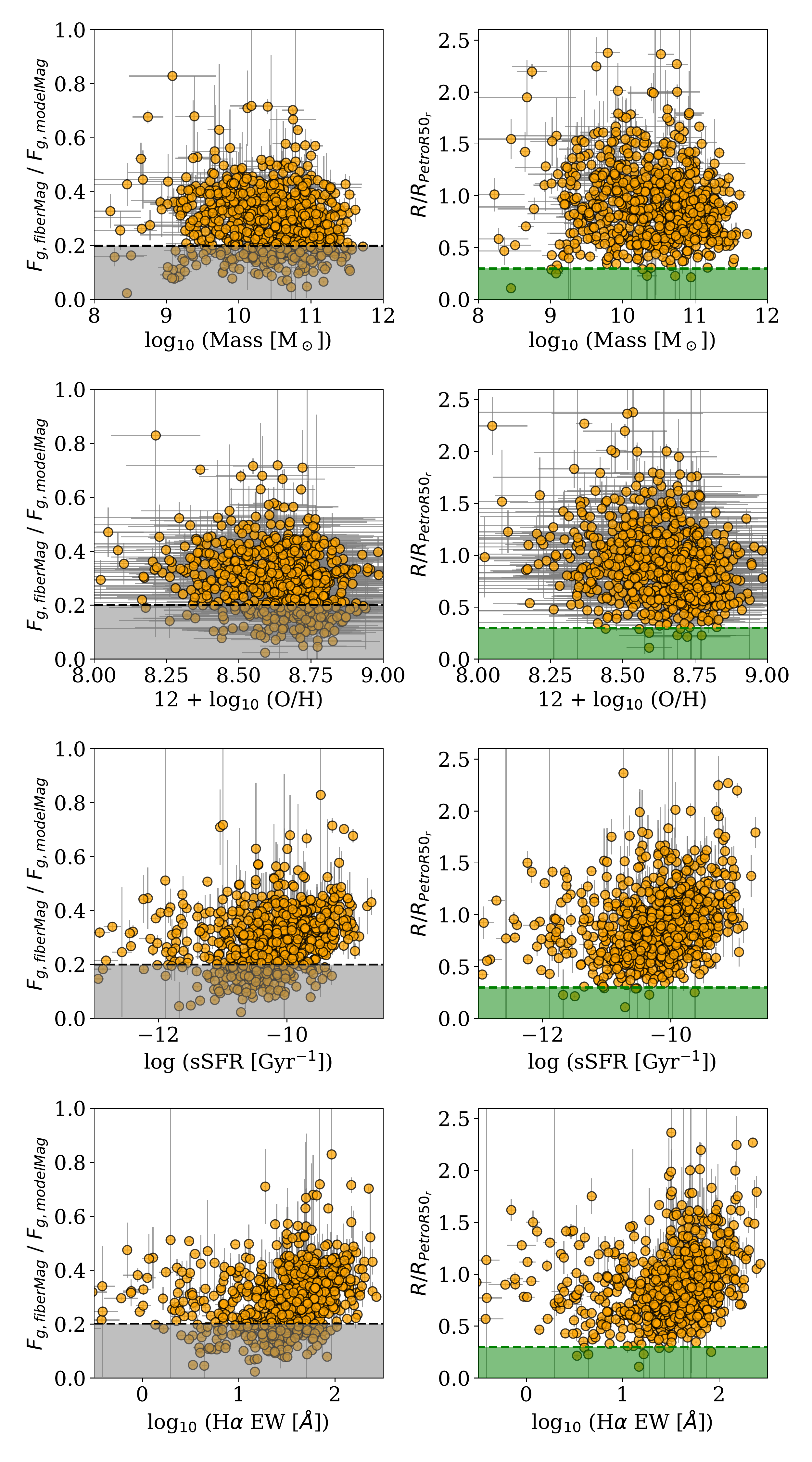}
\caption{$g$-band fraction (left column) and galaxy radius covered by the fiber in R$_{50}$ units (right column) as a function of the main galaxy parameters used in this work: stellar mass, oxygen abundance, specific SFR, and H$\alpha$ equivalent width. The shadowed regions represent the excluded region under each criterion. This Figure clearly shows that under one criterion there are less objects excluded, and the tendency to exclude objects with higher/lower values for each galaxy parameter.}
\label{fig:hostbias}
\end{figure}

\begin{figure}[!t]
\centering
\includegraphics[trim=0cm 0cm 0cm 0cm, clip=true,width=0.9\columnwidth]{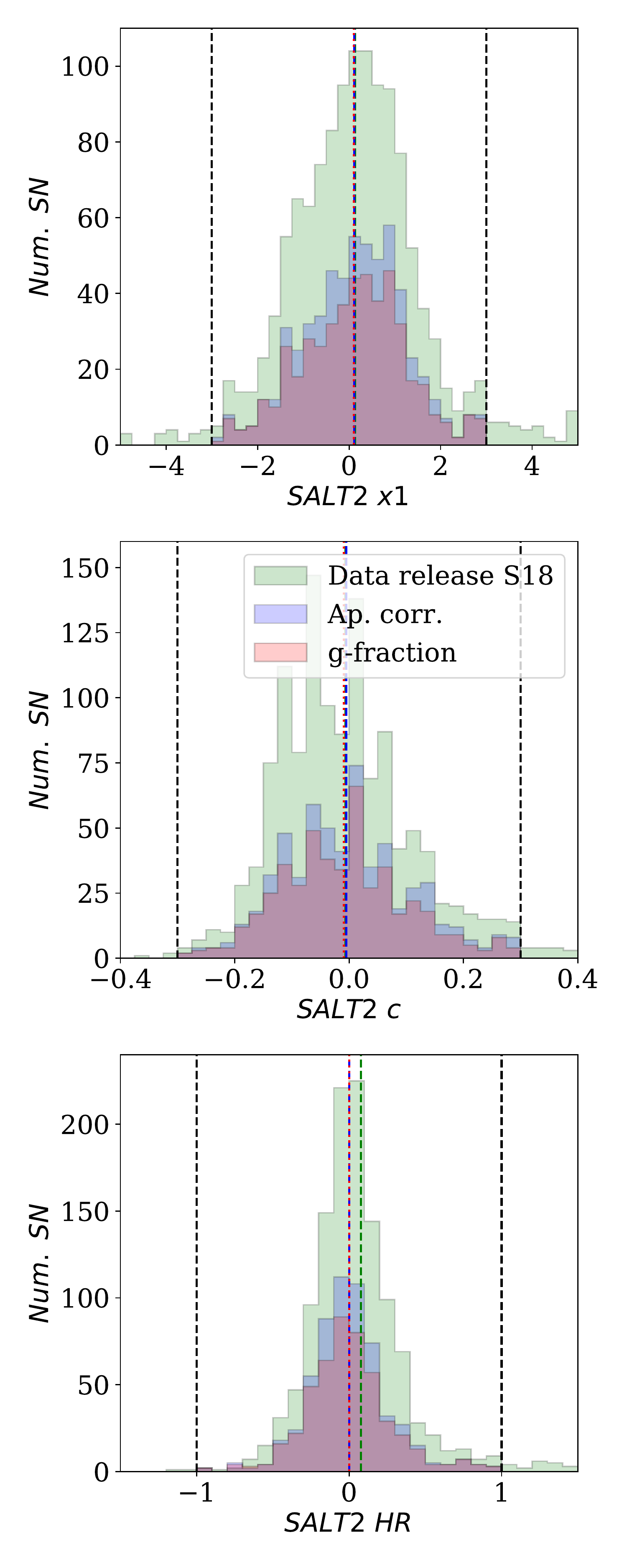}
\caption{$x_1$ and $c$ light-curve parameter and Hubble residual  distributions of the full SDSS-II Data Release (green) compared to those of the samples after applying the fiber radius in R$_{50}$ units (blue) and the $g$-band fraction (in red) cuts used here. Vertical colored lines represent the mean of the samples, and black { dashed} lines define the quality selection range.}
\label{fig:lcpars}
\end{figure}


\section{The effect of aperture corrections on derived properties} \label{sec:eff}

In this section we focus on the  
effect that aperture corrections have on the physical parameters of SN host galaxies for the three different cases of interest: (i) Case A: observed fiber spectra; (ii) Case B: observed fiber spectra scaled to the global galaxy photometry; and (iii) Case C: observed fiber spectra corrected for aperture effects.

The first row of Figure~\ref{fig:mass-relations} shows the SFR -- $M_\star$ relation for Case A, B, and C, from left to right. In each panel, we performed a linear fit only to the star-forming galaxies (the main sequence of this relation). In the lower panel of Case C we show the difference between the Case A/B fit and the Case C fit. From the differences in the linear fits we found that at a fixed stellar mass the SFR values in case C are on average higher than in cases A and B across the stellar mass range between 10$^{8.5}$ and \mbox{10$^{12.0}$ M$_\odot$}. Most importantly, the residuals show a trend with stellar mass implying a mass-dependent bias in the inferred SFR without considering aperture effects: these differences are higher at low stellar masses (0.45 and 0.20 dex for A and B linear fits, respectively) than at the higher end (0.18 and 0.06 dex for A and B, respectively). Therefore, regardless of whether the H$\alpha$ flux has been scaled or not, if it is not corrected for aperture effects, the SFR value is underestimated in the entire range of stellar masses between 10$^{8.5}$ and 10$^{12.0}$ M$_\odot$.

In the central row of the same Figure~\ref{fig:mass-relations} we repeated the same analysis, this time for the O/H -- M$_\star$ relation. 
From the differences in the linear fits shown in the lower panel, we found that the aperture corrected values of O/H at a fixed stellar mass are slightly lower than the values obtained in cases A and B for the stellar mass range between 10$^{8.5}$ and 10$^{12.0}$ M$_\odot$. These differences are small in the whole stellar mass range considered (lower than 0.1 dex), as expected for intrinsic parameters.

In the bottom row of Figure~\ref{fig:mass-relations} we show the sSFR -- M$_\star$ relation in each case. In addition, for case C we show the linear fit of the sSFR values in the fiber (note that it is not the same value as in Case A since the stellar masses considered for case C are scaled). Considering these differences in the fits we found that on average at a fixed stellar mass the sSFR values in case C are higher than in cases A and B across the entire stellar mass range between 10$^{8.5}$ and 10$^{12.0}$ M$_\odot$. Differences are higher at low stellar masses (0.45 and 0.30 dex for A and B, respectively) than at high stellar masses (0.18 and 0.06 dex for A and B, respectively). When we compared the sSFR in case C (using only the scaled stellar mass) for H$\alpha$ flux both corrected and uncorrected for aperture, we found that the difference between them increases with the stellar mass (from 0.30 dex at 10$^{8.5}$ M$_\odot$ to 0.54 dex at 10$^{12.0}$ M$_\odot$).

In Fig.~\ref{fig:hist-diff-sfr} we show the histogram of the differences between Case C and A (aperture-corrrected vs. observed), as well as Case C and B (aperture-corrrected vs. scaled) for the SFR, O/H and sSFR. In addition, we also show the difference in A$_v$ between case C and case A. When aperture corrections are not taken into account, the SFR value is underestimated by $\sim$0.44 dex on average when comparing cases C and A, and $\sim$0.13 dex when comparing cases C and B. The average difference of $\Delta$(12+log(O/H)) is $\sim$-0.025 dex in both cases (C-A and \mbox{C-B}), so aperture-corrected abundances are almost always lower. When we considered the differences in the sSFR values between cases C with A and C with B, we obtain a difference of $\sim$0.13 dex in both cases. Taking into account the galaxies of case C, when we compared the sSFR values with the SFR measurement corrected and uncorrected for aperture effects we recovered the average difference of $\sim$0.44 dex found in the upper-left panel. Finally, we found that the median of the difference between the aperture-corrected and the uncorrected A$_v$ ($\Delta A_{v}$) is -0.16 mag.

\subsection{Comparison to fiber fraction galaxy coverage}

\cite{2005PASP..117..227K} found that for an emission-line metallicity measurement to be representative of the global value, the spectrum should contain $>$20\% of the host-galaxy $g$-band flux; Moreover \cite{2008ApJ...681.1183K} claimed that this $g$-band flux fraction had to be $>$30\% for galaxies with M $>$ 10$^{10}$ M$_\sun$. In either case, this translates to excluding from any metallicity analysis all those galaxies with the $g$-band flux relatively less concentrated or galaxies larger in size.

Thus, the fraction of the total host-galaxy light covered by the fiber is an important quantity to consider in this analysis. Fortunately, spectra obtained as a part of the SDSS survey have fiber and model magnitudes associated with each galaxy, so we can easily compute the observed light fraction in the spectra, and compare the resulting sample with the derived in our approach, which only depends on the Petrosian R$_{50}$ parameter.

In the left panel Figure \ref{fig:apercor2}, we examine the $g$-band fiber fraction for the initial 732 galaxies before applying our aperture corrections, as a function of the total $g$-band of galaxies. We see a clear relation between these two parameters in the direction of fainter galaxies, which in turn are also smaller, to have a larger area covered by the fiber. The criteria of selecting galaxies with $g$-band fiber coverage larger than 20\% would have excluded 159 galaxies, shown under the grey region in this panel. In the central panel we compare the two criteria, the $g$-band light fraction and the fiber radius in R$_{50}$ units. Despite the tight relation between the two parameters, the different criteria to discard galaxies provide samples of significantly different sizes: from the initial 732 galaxies, we kept 573 and 724, respectively. Finally, in the right panel we show the redshift dependence of the fiber radius in R$_{50}$ units. It can clearly be seen that those 8 galaxies in the exclusion zone are both at the lower redshift end and with very low fiber radii in R$_{50}$ units, so large in size. We also marked with black edges those galaxies that pass our criterion for aperture corrections but where the fiber did not cover up 20\% of the $g$-band light, and would have been excluded following that other criterion.

So, we have here shown that under our approach we are able to both (a) correct the measured host galaxy parameters for aperture effects, and (b) keep a larger number of objects for the study of SN host galaxy correlations.

\subsection{Biases on host galaxy parameters}\label{sec:bias}

Here we also show that our sample selection based on the availability of aperture corrections not only maximizes the number of objects for the analysis, but also avoids the introduction of biases in the host galaxy distributions. 

Previous works similar to ours have followed the $g$-band fraction criterion (e.g. \citealt{2011ApJ...743..172D}). For instance, one of the most recent, \cite{2016ApJ...821..115W} found little correlation between g-band fiber fraction and gas-phase metallicity, indicating aperture-corrections were not needed.

In Figure \ref{fig:hostbias} we present the relation between the $g$-band fiber galaxy fraction and the fiber coverage in R$_{50}$ units with the stellar mass, oxygen abundance, specific SFR, and H$\alpha$ equivalent width. While it is evident that the few objects that are excluded using the R$_{50}$ criterion for aperture corrections correspond to outliers, we demonstrate how the $g$-band fraction selection tends to remove from the sample large mass, high metallicity, low sSFR and low H$\alpha$ equivalent width galaxies. 

Summarizing, the approach followed in this work permits to use a significant larger fraction of objects than other approaches, corrects for aperture effects, and avoids selection biases in host galaxy distributions.


\begin{figure*}[t]
\includegraphics[trim=0.8cm 0cm 0.7cm 0cm, clip=true,width=\textwidth]{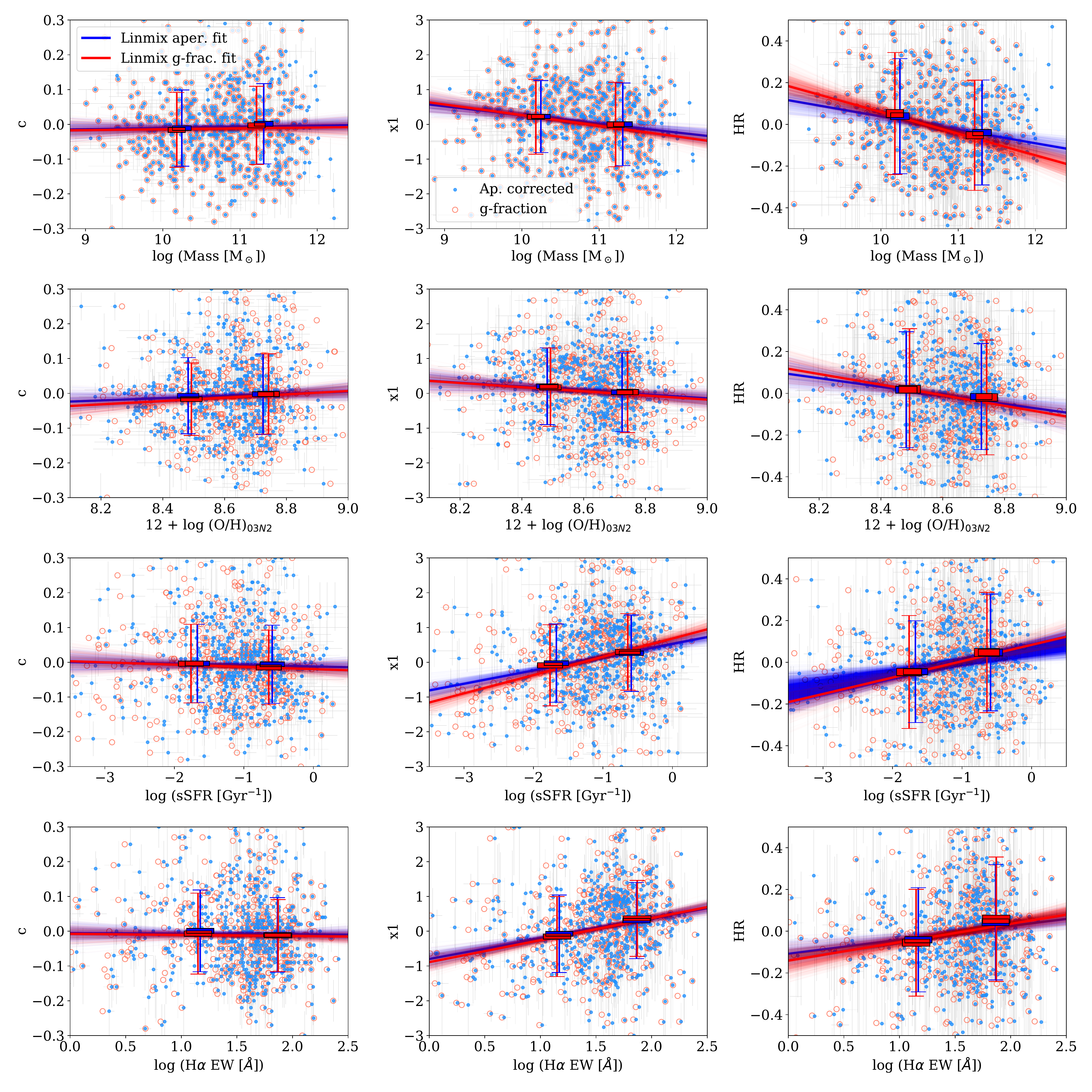}
\caption{$x_1$, $c$ and Hubble residuals correlations with host galaxy parameters: stellar mass, oxygen abundance, specific SFR, and H$\alpha$ equivalent width, for the aperture-corrected sample (in blue) and the $g$-band fraction sample (in red). Solid transparent lines are the 10,000 individual slopes found with the {\sc LINMIX} MCMC sampler, with the average slope as a solid line on top. Overplotted there are the average and  standard deviation of the variable on the Y-axis in each of the two high/low bins in the X-axis variable, which has been divided by using the median X-value as the split point. The height of the box represent the error of the mean in each bin.}
\label{fig:hostrel}
\end{figure*}

\section{SN light-curve parameters}\label{sec:sifto}

\begin{table*}[!t]
\caption{Results of the {\sc LINMIX} MCMC slopes, and difference (steps) between the high and low bin.}
\begin{center}
\begin{tabular}{llcccccc}
\hline
         &            &            c     & $\sigma$ &          x1      & $\sigma$ &          HR      & $\sigma$ \\ 
\hline
\multicolumn{8}{c}{log (Mass [M$_\odot$])}\\
\hline
Ap. cor. & slope   &  0.004$\pm$0.007 & 0.50 & -0.249$\pm$0.071 & 3.49 & -0.064$\pm$0.017 & 3.75 \\
         & step    & -0.012$\pm$0.009 & 1.34 &  0.223$\pm$0.092 & 2.43 &  0.078$\pm$0.022 & 3.60 \\
g-frac   & slope   &  0.003$\pm$0.009 & 0.34 & -0.299$\pm$0.085 & 3.53 & -0.104$\pm$0.022 & 4.73 \\
         & step    & -0.011$\pm$0.010 & 1.11 &  0.239$\pm$0.107 & 2.24 &  0.097$\pm$0.026 & 3.78 \\

\hline
\multicolumn{8}{c}{12 + log (O/H)$_{03N2}$}\\
\hline
Ap. cor. & slope   &  0.031$\pm$0.033 & 0.93 & -0.558$\pm$0.328 & 1.70 & -0.182$\pm$0.076 & 2.39 \\
         & step    & -0.004$\pm$0.009 & 0.48 &  0.177$\pm$0.092 & 1.92 &  0.033$\pm$0.022 & 1.49 \\
g-frac   & slope   &  0.053$\pm$0.036 & 1.46 & -0.597$\pm$0.375 & 1.59 & -0.243$\pm$0.082 & 2.96 \\
         & step    & -0.019$\pm$0.010 & 1.86 &  0.153$\pm$0.105 & 1.46 &  0.051$\pm$0.026 & 1.96 \\

\hline
\multicolumn{8}{c}{log (sSFR [yr$^{-1}$])}\\
\hline
Ap. cor. & slope   & -0.004$\pm$0.007 & 0.60 &  0.383$\pm$0.069 & 5.55 &  0.056$\pm$0.015 & 3.76 \\
         & step    &  0.002$\pm$0.009 & 0.25 & -0.284$\pm$0.092 & 3.09 & -0.092$\pm$0.022 & 4.24 \\
g-frac   & slope   & -0.007$\pm$0.008 & 0.81 &  0.520$\pm$0.082 & 6.30 &  0.077$\pm$0.019 & 4.04 \\
         & step    &  0.014$\pm$0.010 & 1.42 & -0.328$\pm$0.104 & 3.14 & -0.093$\pm$0.026 & 3.59 \\
\hline
\multicolumn{8}{c}{log (H$\alpha$ EW [$\AA$])}\\
\hline
Ap. cor. & slope   & -0.002$\pm$0.009 & 0.26 &  0.582$\pm$0.089 & 6.56 &  0.073$\pm$0.020 & 3.64 \\
         & step    &  0.011$\pm$0.009 & 1.15 & -0.385$\pm$0.091 & 4.21 & -0.086$\pm$0.022 & 3.92 \\
g-frac   & slope   & -0.003$\pm$0.009 & 0.36 &  0.625$\pm$0.088 & 7.14 &  0.093$\pm$0.023 & 4.10 \\
         & step    &  0.003$\pm$0.010 & 0.34 & -0.522$\pm$0.104 & 5.03 & -0.107$\pm$0.026 & 4.19 \\
\hline
\end{tabular}
\end{center}
\label{tab:res}
\end{table*}

To study the intrinsic relationship between the properties of a SN~Ia and its host, we use the light-curve parameters as published in the SDSS-II/SN Data Release \citep{2018PASP..130f4002S}, which were obtained using the SALT2 template implemented in the publicly available Supernova Analyzer package (SNANA; \citealt{2009PASP..121.1028K}).

In the SALT2 model 4 parameters, namely the epoch of maximum brightness in the $B$-band ($t_{0}$), the color of the supernova ($c$) at peak, the $x_1$ parameter (related to the stretch of the {\lc}), and the $x_0$ normalization factor (from which one can obtain the apparent magnitude at maximum brightness in the $B$ band; $m_{B}$) are determined from the fit to the multiband light curve. Standardized magnitudes are obtained using the Tripp equation,
\begin{equation}
\mu_{SALT2}= -\log_{10} x_0 -M_x +\alpha x_1 - \beta c \, ,
\end{equation}
where $M$, $\alpha$ and $\beta$ are nuisance parameters resulting from minimizing the difference between $\mu_{\rm SALT2}$ and the distance modulus assuming a fiducial cosmology. Since our goal is not measure the best cosmology from our data, but look for systematic effects with aperture-corrected host galaxy parameters and correlate the residuals from the distances measured with SNe Ia to that cosmology, in our case we used the $\mu_{\rm SALT2}$ reported values in \citep{2018PASP..130f4002S} for the nuisance parameters ($M, \alpha, \beta$)=\mbox{(-29.967, 0.187$\pm$0.009, 2.89$\pm$0.09)}, and calculated Hubble residuals by subtracting the distance modulus of a flat cosmology from them ($HR = \mu_{\rm SALT2} - \mu_{\rm cosmo}$). To assure a robust sample of SNe Ia unaffected by peculiar objects, we discard SNe Ia with a SALT2 probability of being a SNIa (SNANA parameter {\sc fitprob}) $<$1\%. In addition, we remove all SNe with extreme light-curve parameter values following standard selection cuts for {\salttwo}, so the allowed ranges are set to $-0.3<c<0.3$, $-3.0<x1<3.0$, and -1.0$<HR<$1.0 \citep{2018PASP..130f4002S}. These cuts were designed to remove objects with peculiar or badly constrained {\lc}s. From our list of 724 and 573 SNe Ia with measured host galaxy parameters in the aperture-corrected and the g-band fraction samples, we are left with a final sample of 589 and 475 SNe Ia, respectively.

Figure \ref{fig:lcpars} shows the distribution of SALT2 $c$ and $x_1$ parameters and the Hubble residuals for the full SNe Ia in the SDSS-II/SN DR, and similar distributions for our final sample. All distributions have equivalent shape and statistics ($\left<x1\right>=0.11$, $\sigma_{x1}=1.12$; $\left<c\right>=-0.01$, $\sigma_{c}=0.11$; $\left<HR\right>=0.00$, $\sigma_{HR}=0.27$), but it is evident how the sample decreased from the initial to the aperture-corrected, and even more to the $g$-fraction sample.


\section{Results and discussion}\label{sec:results}

We present SNIa light-curve parameters and Hubble residual correlations with host galaxy parameters in Figure \ref{fig:hostrel}, separately for the 589 objects in the aperture-corrected (our {\it fiducial} sample, as blue dots) and the 475 objects in the $g$-fraction (the {\it literature approach} sample, in red circles) samples. We performed two kind of analyses: (i) 'step' functions: we divided the sample in two bins of the same size using the independent variable in the X-axis to then measure the mean, mean error, and standard deviation of the dependant variable in the Y-axis in each of the two bins; (ii) we performed a linear fit with {\sc LINMIX}\footnote{\href{https://github.com/jmeyers314/linmix.git}{https://github.com/jmeyers314/linmix.git}} \citep{2007ApJ...665.1489K}, which is more robust than a simple least squares linear fit because it takes into account the errors in both \mbox{X-axis} and \mbox{Y-axis} variables, and also because it computes a probability function for the dataset using a Markov-Chain Monte Carlo (MCMC) algorithm. A summary of all slopes and steps found are listed in Table \ref{tab:res}. Overplotted in each panel of Figure \ref{fig:hostrel}, we show the best fit from {\sc LINMIX} (thick solid line; plus the 10,000 minimizations with transparency). 

\subsection{Recovering mass step}

We start testing whether the Hubble residuals of our resulting samples conserve a dependence with the host galaxy mass as reported by several works, including those using SDSS-II/SN data. As noted above, the host mass is not a parameter for which we have aperture corrections available, therefore we focus on how different the dependence is in the two final samples due to possible bias introduced by the selection criterion.

In the first row of Figure \ref{fig:hostrel} we show the light curve parameters $c$ and $x1$, together with the Hubble residuals as a function of the (scaled; case B) mass, for the SNe Ia in the aperture-corrected (blue) and $g$-fraction (red) samples. While we find a low significance ($<$1.3$\sigma$) mass step in the color parameter, we find a 2-4$\sigma$ step in the $x1$ and Hubble residuals. The mass step is higher and more significant in the $g$-band fraction sample than in the aperture-corrected, which in principle contains a wider and more complete set of host galaxy parameters. In particular, we showed in section \ref{sec:bias} that removing objects with small $g$-band fraction tend to get rid of higher mass galaxies. It is evident in the upper-right panel of Figure \ref{fig:hostrel}, Hubble residuals vs. mass, that most of the blue points at larger masses are not included in the $g$-band sample, and most of them have positive Hubble residuals. Therefore, the g-band fraction criterion introduces a selection bias since it preferentially tends to include galaxies whose light is more concentrated (elliptical morphology, early-type). Since these galaxies on average host SNe with particular properties, the net effect is to artificially increase the height of the mass step.

\begin{figure*}[!t]
\includegraphics[trim=0cm 0cm 0cm 0cm, clip=true,width=\textwidth]{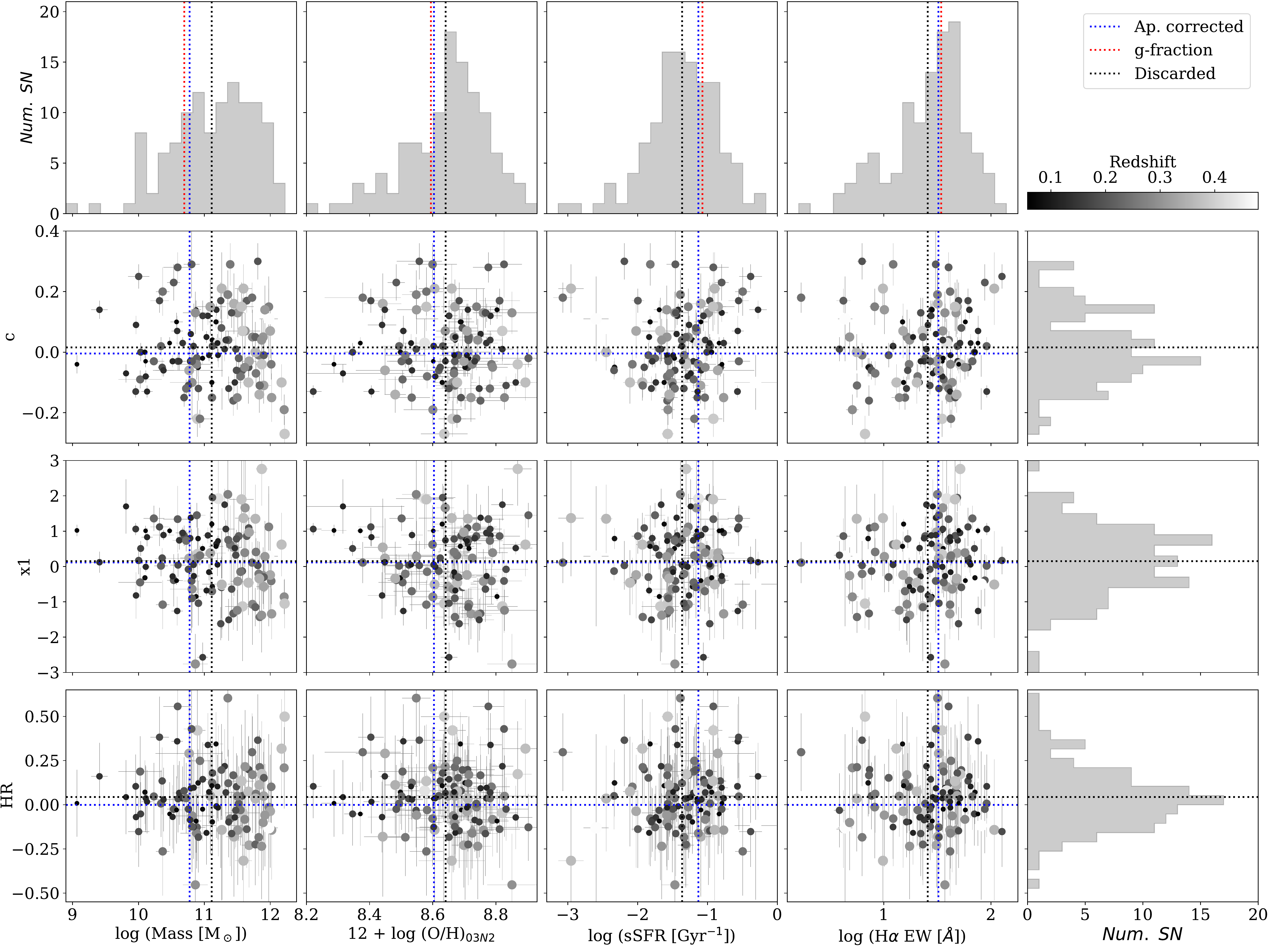}
\caption{SNIa light-curve $x1$ and $c$ parameters and Hubble residual dependences on stellar mass, oxygen abundance, specific SFR, and H$\alpha$ equivalent width of the 114 objects that are discarded using the $g$-band fraction criterion. Black lines represent the mean of the distribution, and red and blue lines represent the mean of the $g$-band fraction and aperture-corrected samples, respectively. Regarding light-curve parameters and Hubble residuals, while the three original distributions are centered at zero, these discarded objects have positive averages in the three distributions, leaving on average {\it more negative} biased distributions in the $g$-band fraction sample. Moreover, discarded objects have on average larger stellar mass, lower H$\alpha$ EW, lower sSFR, and higher metallicity with respect the aperture-corrected sample, so the $g$-band fraction criterion leaves as well a sample of SNIa and host galaxies that are biased towards the other side of the distributions.}
\label{fig:bias}
\end{figure*}

In Figure \ref{fig:bias} we demonstrate these two (and other) biases, with distributions of the SNIa light-curve parameters, Hubble residuals, and host galaxy parameters for the 114 objects that are present in the aperture-corrected sample but that did not pass the $g$-band fraction criterion. These objects have slightly larger $x1$ ($\sim$0.15), $c$ ($\sim$0.02), and Hubble residuals ($\sim$0.04 mag) than the average value in the aperture-corrected sample. Moreover, they show an average stellar mass that is $\sim$0.34 dex larger then the average mass of the aperture-corrected sample. The net effect of the g-band fraction criterion is discarding intrinsically faint SNe Ia in higher-mass galaxies, thus artificially increasing the height of the mass step.

\subsection{SN light-curve dependencies on corrected host galaxy parameters}

In the bottom three rows of Figure \ref{fig:hostrel} we present the dependence of light-curve parameters and Hubble residuals with host galaxy oxygen abundance, specific SFR, and H$\alpha$ equivalent width. Our assumption along this work is that the aperture-corrected sample is unbiased with respect to the selection performed by removing all galaxies with fiber fractions smaller than a certain limit. The relations found in the aperture-corrected sample are therefore the {\it true underlying correlations} of $x1$, $c$, and the Hubble residuals, with the  late-type star-forming host galaxy environmental parameters.

We find a significant trend between the $x1$ parameter and the Hubble residual and both the specific SFR and H$\alpha$ equivalent width (3-6$\sigma$), in the sense of wider and intrinsically faint (positive HR) SNe Ia occurring in galaxies with higher specific SFR and H$\alpha$ EW, consistent with previous findings (e.g. \citealt{2006ApJ...648..868S,2010MNRAS.406..782S}). On the other hand, the $c$ step are of the order of 1$\sigma$, while the slope is insignificant. Specific SFR is measured dividing the aperture-corrected H$\alpha$ flux (case C) by the scaled stellar mass (case B). It therefore represents the efficiency of a galaxy to form stars normalized by the total mass of the stars present. On the other hand, H$\alpha$ EW is the ratio of the H$\alpha$ flux over the flux of the continuum, so it provides a measurement of how intense is the current star formation compared to the past SFR. Although similar, the information these two indicators provide is not exactly the same. Although these correlations are present in both samples, in general (in all cases except in the step of the sSFR), they are marginally increased in the $g$-band fraction sample. Thus, the sample selection also introduces a bias in the sample towards increasing the significance of these correlations.

Regarding oxygen abundance, we find insignificant relations with the $c$ parameter ($<1\sigma$), in the sense of bluer SNe Ia occurring in higher metallicity galaxies, that is marginally increased in the aperture-corrected sample. Lower significance (2-3$\sigma$) trends are found with $x1$ and Hubble residuals, where negative $x1$ and negative $HR$ tend to occur in high-metallicity galaxies. In particular, we find a metallicity step in the Hubble residual with a significance of $\gtrsim$2$\sigma$. All these are in line with previous results in the literature \citep{2011ApJ...743..172D,2018MNRAS.476..307M}. Similarly, the $g$-band fraction results have higher significance with Hubble residuals, pointing to a selection bias.

Again, in Figure \ref{fig:bias} where the distribution of these other host galaxy parameters are shown for the 114 SNe Ia in the sample discarded using the $g$-band fraction criterion, together with the correlations between them and the light-curve parameters and Hubble residuals. Discarded objects have on average an oxygen abundance 0.04 dex higher, a $\log_{10} sSFR = -0.25$ dex lower, and $\log_{10} {\rm H\alpha EW} = -0.12$ dex lower, than both the full aperture-corrected and the $g$-band fraction selected samples. These differences explain the varying significance between the two main samples in all relations. Moreover, we note that for the aperture-corrected sample the step significance with environmental parameters is sorted from higher to lower as (see Table \ref{tab:res}):
\begin{equation}
{\rm sSFR} \rightarrow {\rm H\alpha EW} \rightarrow {\rm Mass} \rightarrow {\rm O/H}, 
\end{equation}
while for the g-fraction sample is,
\begin{equation}
{\rm H\alpha EW} \rightarrow {\rm Mass} \rightarrow  {\rm sSFR} \rightarrow {\rm O/H}.
\end{equation}
Besides the difference in the sorted parameters, we note how in the fiducial sample both the sSFR and the H$\alpha$EW show a greater significance than the galaxy stellar mass. As detailed above, although sSFR and H$\alpha$ equivalent width are not exactly providing the same information about the underlying stellar populations, they both are proxies for the stellar population age. Therefore, our findings indicate that, once getting rid of the selection biases, the stellar population age seems to have a greater impact in the standardization of SNe Ia than the stellar mass (as pointed out by \citealt{2019ApJ...874...32R}). Historically, mass has been widely used in cosmological analyses to improve the standardization, and the mass step is now a common term added in the Tripp equation \citep{2014A&A...568A..22B,2019ApJ...874..150B}. This is partially explained by the simplicity of its measurement; several works have found that with just a few photometric points it can be constrained with high accuracy \citep{2013ApJ...770..107C,2021arXiv210208980H}. Stellar age is more difficult to constrain, specially with photometric data alone. It is notably degenerate with stellar metalicity and dust reddening, and it is better determined from spectroscopy or with complementary UV and NIR data.

Similarly to e.g. \cite{2018A&A...615A..68R}, we studied whether the information encapsulated in the sSFR can be recovered with the stellar mass alone, by performing the following test: for each of the four host galaxy parameters in Figure \ref{fig:hostrel} we corrected the higher bin of the Hubble residual relation by the height of the measured step, and looked again for the correlations. While after correcting for the mass step, we still find a $>$2$\sigma$ significance in the sSFR and H$\alpha$EW steps, the significance of the mass step is $<$2$\sigma$ once either the sSFR or the H$\alpha$EW step is corrected for. This may indicate that sSFR and H$\alpha$ equivalent width, the parameters more related to the age of the underlying stellar populations, do a better job when improving SN~Ia standardization than stellar mass. In the case of metallicity, when its step is corrected the significance of all other three steps is reduced by 0.5-0.7$\sigma$, and remain significant at the 3$\sigma$ level. Also interesting is that, when any of the other three step is corrected for, the remaining metallicity step is insignificant ($<$0.2$\sigma$), indicating that the information captured by the metallicity is already included in either the mass or the stellar age.

\begin{figure}[t]
\includegraphics[trim=0cm 0cm 0cm 0cm, clip=true,width=\columnwidth]{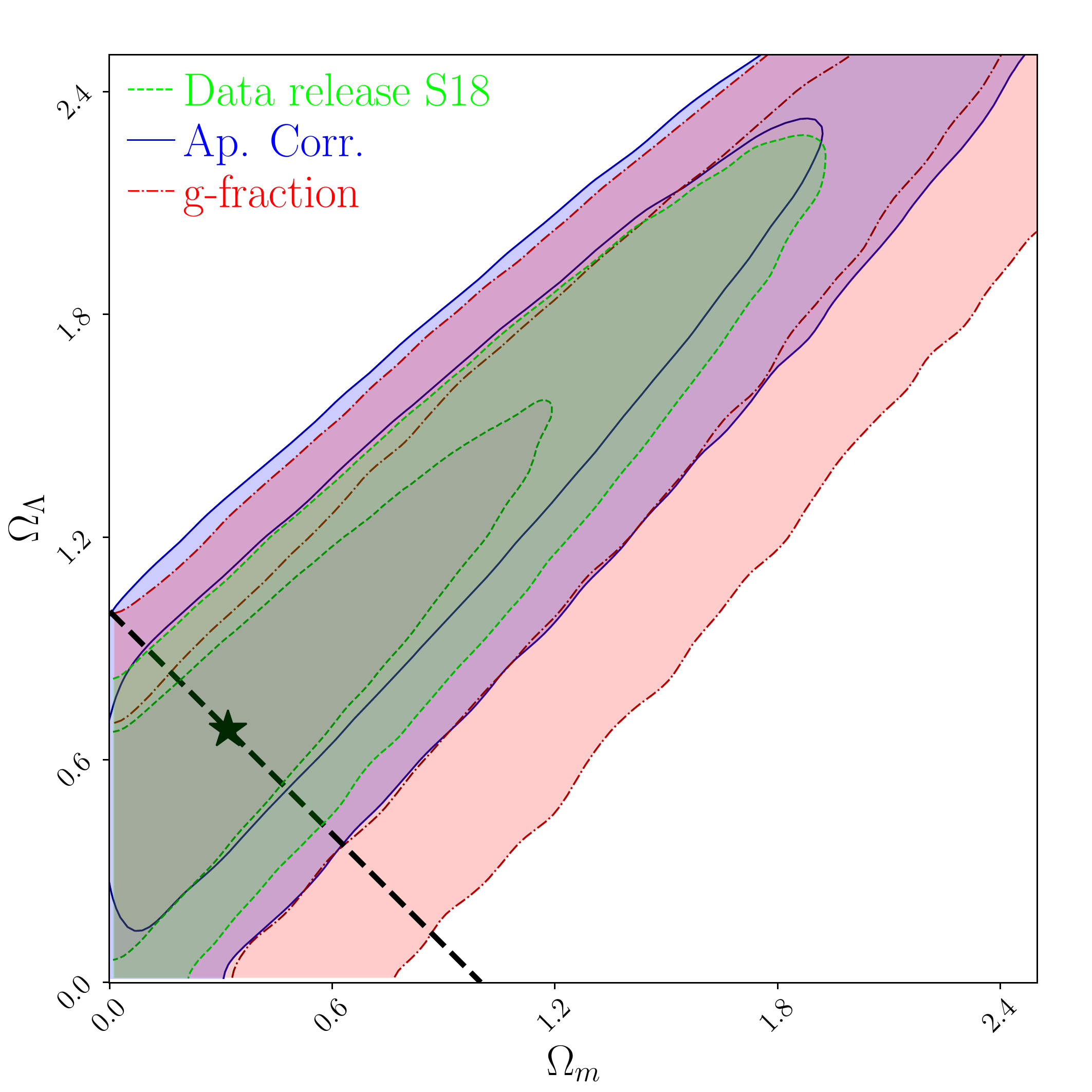}
\includegraphics[trim=0cm 0cm 0cm 0cm, clip=true,width=\columnwidth]{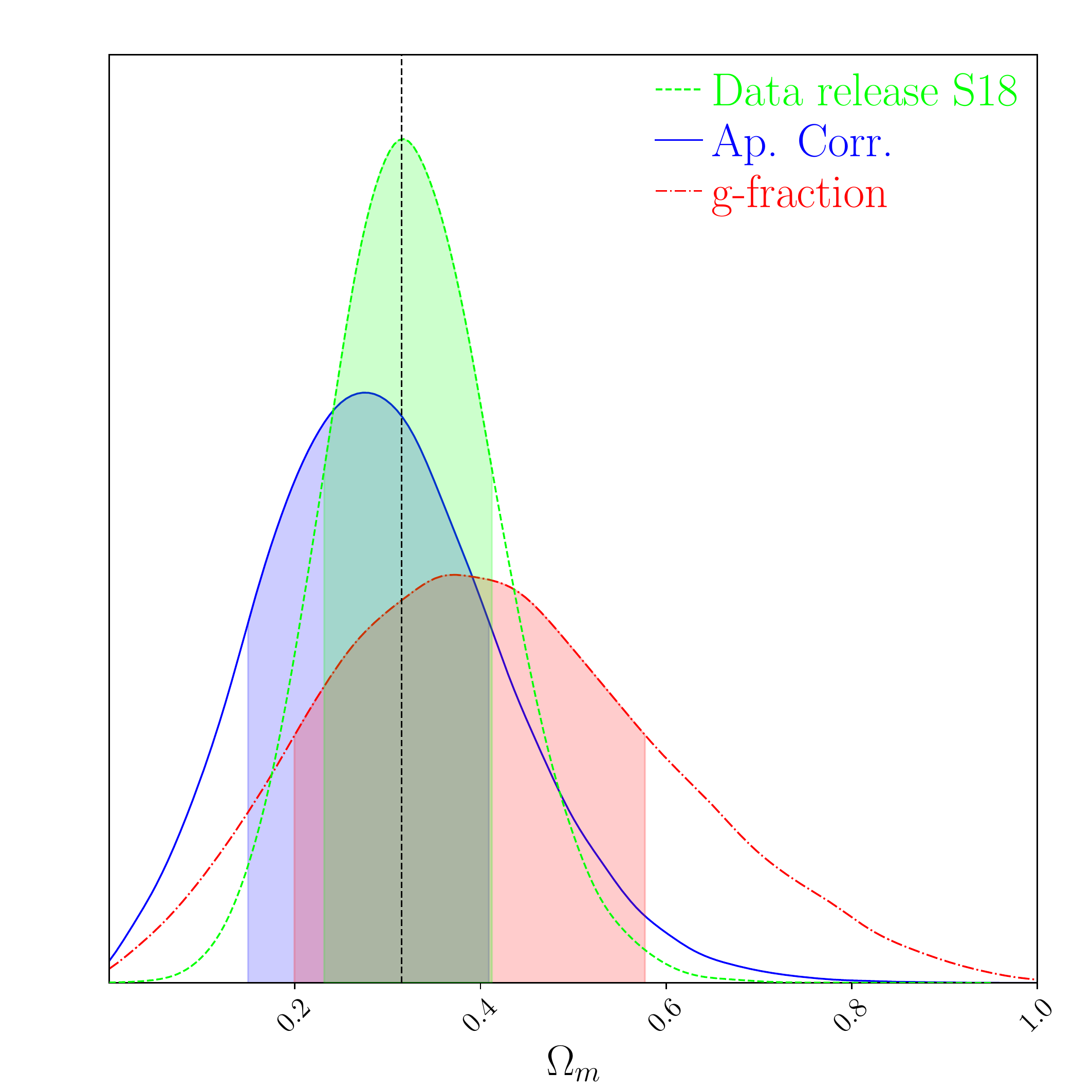}
\caption{{\it Top:} The cosmological constraints on $\Omega_m$,$\Omega_\Lambda$ when considering the aperture corrected sample (blue), compared to the $g$-band fraction sample (red) and the entire SDSS sample (green). The best-fit value for the SDSS dataset, assuming a Flat $\Lambda$CDM model (plotted as a black dashed line) is shown as a black star. Differences in the contours are primarily due to different host galaxy selection. {\it Bottom:} 1D, marginalised contours for $\Omega_m$ are shown for the three samples, once a flat $\Lambda$CDM cosmology is assumed.}
\label{fig:cosmo}
\end{figure}

\subsection{Cosmological parameters determination}

Table~\ref{tab:res} shows the relations between the Hubble diagram residuals and galaxy properties we found for both the aperture-corrected sample and $g$-band fraction sample. With each sub-sample being prone to different selection effects, we here test the effect that these relationships have on the derived cosmological parameters. 

To determine the cosmological parameters, we replicate the selection criteria of \citet{2018PASP..130f4002S}; restricting our analysis to spectroscopically confirmed events, enforcing $-2<x_1<2$ and $-0.2<c<0.5$ and removing objects with poorly measured $x_1$: $\sigma x_1<2$. This selection criteria reduces the aperture-corrected sample to 190 SNe Ia, compared to 122 for the $g$-band fraction sample and 413 for the SDSS sample. To estimate the distance to each SN we use the best-fit nuisance parameters determined in \citet{2018PASP..130f4002S}: namely $\alpha=0.155$, $\beta=3.17$ and $M_0=-29.967$ and apply the Tripp formula \citep{1998A&A...331..815T}. Based on these distances, we then constrain the cosmological parameters: in particular the matter and dark-energy density ($\Omega_m,\Omega_\Lambda$), marginalising over $H_0$ using an MCMC approach. 

In the upper panel of Figure~\ref{fig:cosmo} we show the resulting cosmological constraints from all 3 samples, where the sizes of the contours clearly increase as the sample size is reduced. Going one step further, and assuming a Flat $\Lambda$CDM model (i.e. $\Omega_\lambda=1-\Omega_m$) we recover best-fit estimates of $\Omega_m=0.316^{+0.095}_{-0.086}$ for the SDSS sample, $\Omega_m=0.27^{+0.14}_{-0.12}$ for the aperture corrected sample, and $\Omega_m=0.36^{+0.36}_{-0.16}$ for the $g$-band fraction sample, as shown in the bottom panel of Figure~\ref{fig:cosmo}. All estimates are consistent at $<1\sigma$ indicating that selection biases, while different for each sample, do not overly bias the cosmological parameters. 

Subtracting off the best-fit model ($\Omega_m=0.316^{+0.095}_{-0.086}$) for each sample, we find a root-mean-square scatter (RMS) of 0.183 for the SDSS sample, 0.167 for the aperture corrected sample and 0.174 for the $g$-band fraction sample. The RMS is largest for the SDSS sample, which probes the largest redshift range, and hence includes the lowest signal-to-noise SNe of all 3 samples. Conversely, the smallest RMS is to be found in the aperture corrected sample, which includes additional low-redshift, and hence higher signal-to-noise events, than the $g$-band fraction sample. This result suggests that selecting events based on fiber-fraction as opposed to aperture corrections, will not only cause additional bias in the sample selection, but will likely degrade the inferred cosmological parameters due to the lost of bright, low-redshift, anchor events. We also note that relative error on $\Omega_m$ is smaller for the aperture corrected sample than the $g$-band fraction sample even after taking into account the larger sample size and increased redshift leverage of the aperture corrected sample. This result may hint at future cosmological surveys improving their constraints on the equation-of-state of dark energy in using aperture corrections when inferring galaxy properties, simply due to the reduced uncertainties due to selection effects in this approach. 





\begin{figure}[t]
\includegraphics[trim=0cm 0cm 0cm 0cm, clip=true,width=\columnwidth]{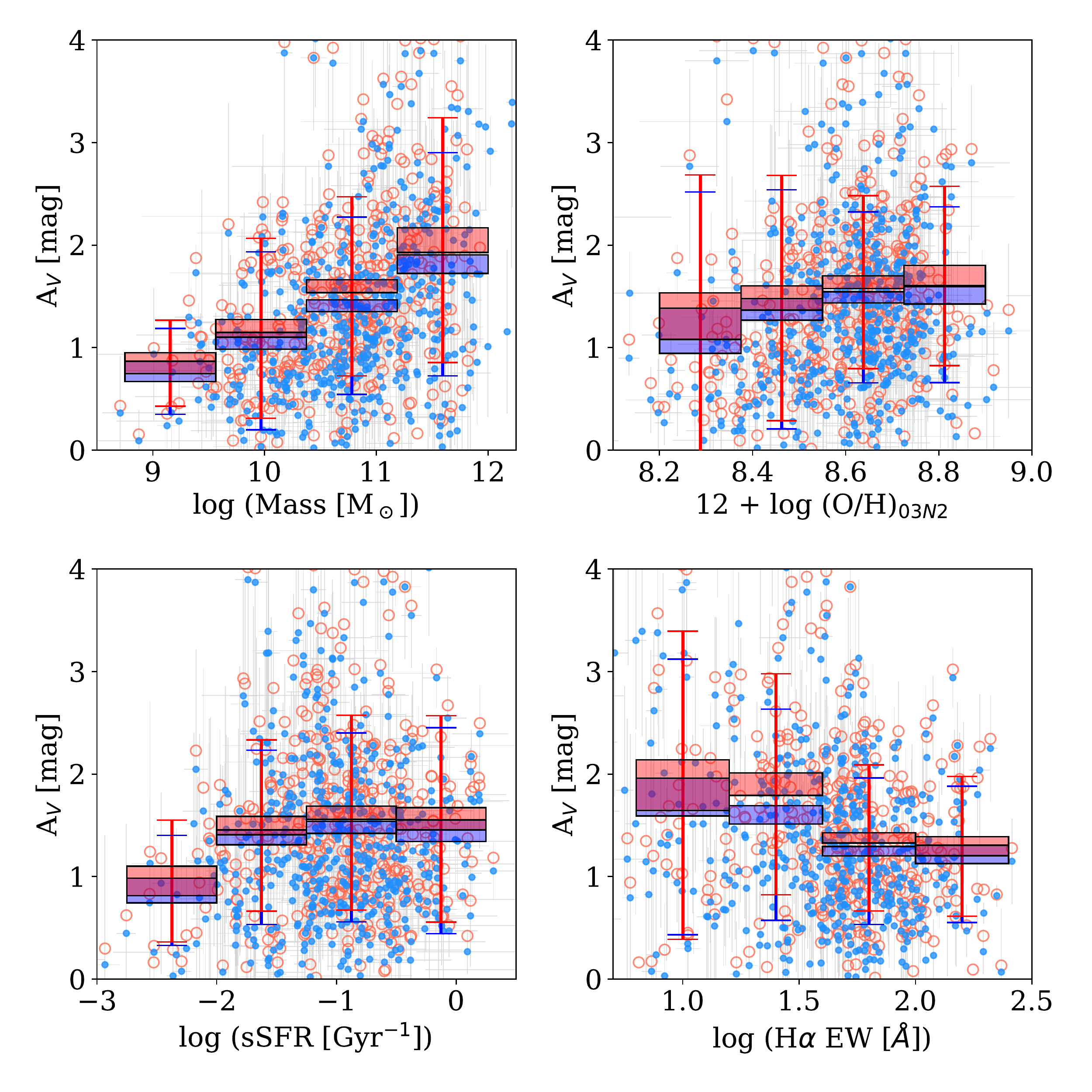}
\caption{Visual extinction as a function of the main host galaxy parameters studied in this work. Blue and red symbols represent the aperture-corrected and the $g$-band fraction samples, respectively. The boxes width represent the error of the mean and the vertical lines represent the scatter in that bin.}
\label{fig:avhost}
\end{figure}

\subsection{Extinction parameter}

We find no significant dependence between any of the light curve parameters or Hubble residuals with the extinction parameter $A_V$. To look further in the dependence of the extinction with host galaxy parameters, and motivated by the recent studies by \cite{2021ApJ...909...26B} and \cite{2021MNRAS.508.4656G}, we present in Figure \ref{fig:avhost} how the measured visual extinction varies as a function of the other host galaxy parameters. As we have shown in Section \ref{sec:aper}, aperture-corrected extinction tends to be lower than extinction measured scaling the spectrum by 0.16 mag on average.

Low-mass galaxies show lower values and with a lower spread in extinction compared to galaxies with higher masses, as previously pointed out by \cite{2010MNRAS.409..421G}. Extinction tends to mildly increase with with oxygen abundance and the dispersion is lower in metal-rich galaxies than in their metal-poorer counterparts. Regarding sSFR and H$\alpha$EW we find different but equivalent behaviours: galaxies with low sSFR and galaxies with large H$\alpha$EW have lower extinction values and with lower spread.

The behaviour in the aperture corrected and $g$-band fraction samples is quite similar, and besides the shift in values we do not find a significant difference among samples. It is out of the scope of this paper a more in depth analysis using different $R_V$ parameters for each SN~Ia or populations (e.g. low/high mass bin), and how this may affect host galaxy environmental dependencies, but it is a topic of on-going research.


\section{Conclusions}\label{sec:disc}

This is the third of a series of papers where we studied how the environment affect SN Ia properties, after presenting a SNIa absolute magnitude dependence on the environmental metallicity \citep{2016MNRAS.462.1281M}, and a revised study of SN Ia light-curve parameters and distances correlations with spectroscopic host galaxy parameters \citep{2018MNRAS.476..307M}. Here, we used the published sample of SNe Ia observed by the SDSS-II SN survey to look for correlations between the SN light-curve parameters and spectroscopic host galaxy parameters from spectra available in SDSS Data Release 16. We used single stellar population synthesis technique to measure the stellar population parameters of the host galaxies and, after removing the stellar contribution, the gas emission parameters. We applied aperture corrections derived from Integral Field Spectroscopy (IFS) provided by the CALIFA survey to the measured parameters, and obtained values that are representative of the whole galaxy getting rid of the different fiber coverage of each galaxy.

Compared to other methods used in previous works, such as the fraction of the total $g$-band flux collected in the fiber, we demonstrate that applying these corrections avoids biasing the sample of galaxies available for the study of SN Ia light-curve and distance correlations with their environmental properties. In particular, the net effect of the $g$-band fraction criterion is discarding intrinsically faint SNe Ia in higher-mass galaxies, thus artificially increasing the height of the mass step. We find equivalent biases in other environmental parameters: high-metallicity, low sSFR, and low H$\alpha$ equivalent width galaxies are similarly removed using the g-fraction criterion, leaving biased samples towards the opposite direction. The order of significance in the Hubble residual dependence between the two samples switches from stellar mass to specific SFR being the most significant in the biased and unbiased samples, respectively.

This novel approach presented here could be performed in the future to a larger and more complete sample of SNe Ia, such as the Dark Energy Survey (DES) SN sample, which has been complemented with fiber-spectroscopy of their host galaxies by the OzDES project \citep{2020MNRAS.496...19L}. Similarly, in a few years the same method could also be applied to the Legacy Survey of Space and Time (LSST) SNe Ia host galaxy spectroscopy that will be collected by dedicated fiber-spectroscopy surveys such as the Dark Energy Science Instrument (DESI; \citealt{2014SPIE.9147E..0SF}), and the Time-Domain Extragalactic Survey (TiDES; \citealt{2019Msngr.175...58S}) within the 4-metre Multi-Object Spectroscopic Telescope (4MOST; \citealt{2019Msngr.175....3D}) consortium. 

Improvements over the corrections presented here would include a complementary IFS project that obtain observations of most/all galaxies hosting SNe discovered by a SN survey. Of course, this would be expensive in time and resources. One has to balance the more precise and higher quality data provided by IFS, and the availability of larger samples of galaxies provided by massive fiber-spectroscopic surveys. Current samples of SN Ia host galaxies observed with IFS such as PISCO \citep{2018ApJ...855..107G} and AMUSING \citep{2016MNRAS.455.4087G} are trying to fill this gap and compile a large sample of objects to allow environmental studies of SNe Ia and their host galaxies with high quality observations.


\begin{acknowledgements}
L.G. acknowledges financial support from the Spanish Ministerio de Ciencia e Innovaci\'on (MCIN), the Agencia Estatal de Investigaci\'on (AEI) 10.13039/501100011033, and the European Social Fund (ESF) "Investing in your future" under the 2019 Ram\'on y Cajal program RYC2019-027683-I and the PID2020-115253GA-I00 HOSTFLOWS project, and from Centro Superior de Investigaciones Cient\'ificas (CSIC) under the PIE project 20215AT016. MS is funded by the European Reearch Council (ERC) under the European Union's Horizon 2020 Research and Innovation program (grant agreement no 759194 - USNAC). SDP is grateful to the Fonds de Recherche du Qu\'ebec - Nature et Technologies. SGG acknowledges support by FCT under Project CRISP PTDC/FIS-AST-31546/2017 and Project~No.~UIDB/00099/2020. SDP, JIP, JMV acknowledge support from the Spanish Ministerio de Econom\'ia y Competitividad under grant PID2019-107408GB-C44, and Junta de Andaluc\'ia Excellence Project P18-FR-2664, and from the State Agency for Research of the Spanish MCIU through the ‘Center of Excellence Severo Ochoa’ award for the Instituto de Astrof\'isica de Andaluc\'ia (SEV-2017-0709). Escrit en part al Bellver, prop del mugr\'o del Tagamanent, sobre la vall de ciment malalta (J. G., center forward).
\end{acknowledgements}

\bibliographystyle{aa}
\bibliography{sdssspdist}

\begin{appendix}

\section{SDSS-II/SNe host matching} \label{sec:sdss}

In general, our matching coincided with the reported in the SDSS-II/SNe DR, however there were a few differences. Eight objects (See Table \ref{tab:badz}) were associated to a galaxy (pass criterion 1), but the SN redshift reported in the DR was different from the redshift of the host galaxy in DR16 (do not pass criterion 2). For those matched objects with available spectrum, we did a visual inspection to ensure that the correct host has been associated with each SN, and to discard bad associations related with morphology and inclination of galaxies. 

For other 16 objects, our preferred associated host galaxy was different from that reported in the DR, due to other reasons (see Table \ref{tab:rescued}): (a) {\it FAINT}: the automatic procedure used in the DR, matched the closest photometric object (in Direction Light Radius, DLR, units), but this object is classified either as a star or a very faint galaxy in SDSS DR16. Visually, other object seemed more reliable as a host galaxy; (b) {\it SPEC}: the matched galaxy in the DR does not have measured spectrum and its redshift was estimated from photometry. We visually checked those cases and assigned a close host with a redshift compatible with that of the SN. (c) {\it MISSING}: no galaxy was assigned in the SN DR, but it was visually clear that a galaxy could be matched. (d) {\it OTHER}: SNe could be matched to two galaxies, and we decided to change the assignment of the DR based on the better agreement in redshifts and positions. We also note that there were twenty of cases where the spectrum was not obtained at the galaxy core, but at the SN position due to the large extent of the galaxies, and these objects are excluded from our analysis.
Additionally, we excluded 9 objects (See Table \ref{tab:qso}) that, although they pass both criteria, had the associated galaxy specturm flagged in DR16 as contaminated by Quasi Stellar Object (QSO; quasars) light, so then no further stellar population analysis could be performed.

We note that a number of host galaxy spectra was also released in the SDSS-II/SNe DR. They were obtained with a number of different telescopes, instruments, and configuration. The aim of those data was to obtain host galaxy redshifts and, although they were successful in meeting their goal, the quality and S/N of the observations was in most cases not optimal for the analysis of the stellar populations. We have visually checked them and determined that spectra of additional 55 galaxies were good enough for measuring their parameters. However, we decided not to include them and keep the analysis of the host galaxy spectra with data taken with a single configuration of the SDSS and BOSS spectrographs.

\begin{table}[!t]
\caption{Host galaxy redshifts used in this work that are different to 3\% to those reported in DR.}
\begin{center}
\begin{tabular}{lcccl}
\hline\hline
SN ID & Type    &z in SN DR & z in this work & \% diff \\
\hline 
 9326 & phot-Ia & 0.3500 &  0.2911   & 16.8 \\
10450 & phot-Ia & 0.2720 &  0.5406   & 98.8 \\
13655 & spec-Ia & 0.2520 &  0.2626   &  4.2 \\
13956 & spec-Ia & 0.2620 &  0.4291   & 63.8 \\
17647 & phot-Ia & 0.4340 &  0.2746   & 36.7 \\
17875 & spec-Ia & 0.2120 &  0.2324   &  9.6 \\
18945 & spec-Ia & 0.2700 &  0.2791   &  3.4 \\
19757 & spec-Ia & 0.4030 &  0.2189   & 45.7 \\
\hline
\end{tabular}
\end{center}
\label{tab:badz}
\end{table}
\begin{table}[!t]\tiny
\caption{Host galaxies changed from the SDSS-II/SNe Survey DR.}
\begin{center}
\begin{tabular}{lccl}
\hline\hline
SN ID & Galaxy in SN DR & Galaxy this work & Reason\\
\hline
1580 & 1237663783143473374 & 1237663783143473373 & SPEC \\
2064 & 1237663457242186614 & 1237663457242186612 & FAINT \\ 
4046 & 1237657191976534239 & 1237657191976534238 & FAINT \\ 
4612 & 1237663783129580009 & 1237663783129580008 & FAINT \\
7051 & 1237663784196767937 & 1237663784196767936 & SPEC \\
8888 & 1237657191444316872 & 1237657191444316871 & FAINT \\
11650 & 1237663784204501159 & 1237663784204501520 & OTHER \\ 
12136 &1237663783656161962 & 1237663783656161959 & FAINT \\
13344 & 1237663543685546296 & 1237663543685546294 & OTHER \\
13903 & 1237666408442298652 & 1237666408442298654 & SPEC \\ 
15103 & 1237663542612656918 & 1237663542612656919 & SPEC \\
15640 & --- & 1237678617401098520 & MISSING \\  
16626 & 1237663783127613910 & 1237663783127613911 & SPEC \\
18945 & --- & 1237657189835866770 &MISSING \\
20345 & --- & 1237657191446741848 &MISSING \\
20578 & 1237663479257170514 & 1237663479257170513 & OTHER \\ 
\hline
\end{tabular}
\end{center}
\label{tab:rescued}
\end{table}
\begin{table}[!t]
\caption{Objects classified as QSOs and removed from the analysis.}
\begin{center}
\begin{tabular}{lc}
\hline\hline
SN ID & Object ID in DR \\
\hline
  5966  & 1237663784741372369 \\
  15272 & 1237663784193360056 \\
  15988 & 1237663784206533181 \\
  18283 & 1237657070090781290 \\
  18782 & 1237657584949789257 \\
  18827 & 1237678437015552153 \\
  19778 & 1237666407364165976 \\
  20245 & 1237663544224842023 \\
  21502 & 1237656906352296084 \\
\hline
\end{tabular}
\end{center}
\label{tab:qso}
\end{table}


\section{STARLIGHT SSP fits} \label{sec:app}

\begin{figure*}[!t]
\centering
\includegraphics*[trim=0cm 0cm 0cm 0cm, clip=true,width=\textwidth]{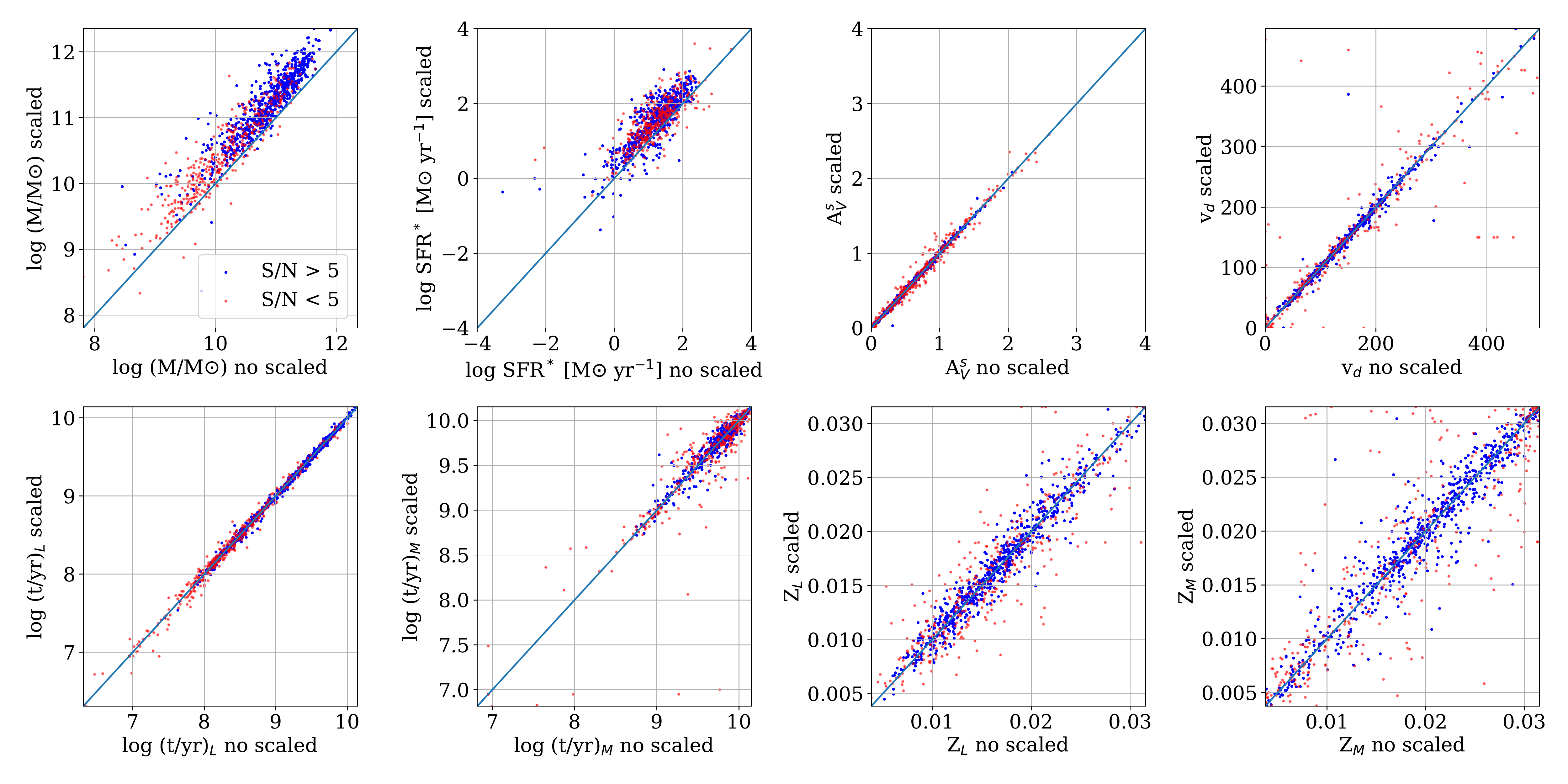}
\caption{Differences between STARLIGHT SSP fit between observed and scaled to the SDSS photometry spectra. Extensive parameters (mass, SFR) are shifted from the diagonal, and all other intensive parameters are very consistent.}
\label{fig:ssp}
\end{figure*}

Comparison of simple stellar population (SSP) fits results using STARLIGHT between observed and scaled spectra. See Figure \ref{fig:ssp} where we show the stellar mass, star formation rate (SFR), visual extinction of the stellar component ($A_V^s$), velocity dispersion ($v_d$), and average stellar age and metallicity luminosity- and mass-weighted.

\section{Measured and corrected parameters} \label{app:tab}

\begin{table*}
\tiny
\caption{Host galaxy parameters measured from the observed spectrum (case A), scaling to the total photometry (case B), and after applying aperture corrections (case C).}
\begin{tabular}{ c| c c | c c c | c c c | c c | c c}
\hline\hline \\[-2ex]
(1) & (2) & (3) & (4) & (5) & (6) & (7) & (8) & (9) & (10) \\[0.5ex]
ID & log(M$_\star$/M$_\odot$)$_{fibre}$ & log(M$_\star$/M$_\odot$)$_{scaled}$ & log(SFR/M$_\odot$ yr$^{-1}$)$_{fibre}$ & log(SFR/M$_\odot$ yr$^{-1}$)$_{scaled}$ & log(SFR/M$_\odot$ yr$^{-1}$)$_{corr}$ & 12+log(O/H)$_{fibre}$ & 12+log(O/H)$_{scaled}$ & 12+log(O/H)$_{corr}$ & A$_{V,\ fib}$ [mag]& A$_{V,\ corr}$ [mag]& EW(H$\alpha$)$_{fib}$ [\AA]& EW(H$\alpha$)$_{corr}$ [\AA]\\
\hline\\[-2ex]
5 & 10.23$\pm$0.10 & 10.54$\pm$0.09 & 0.21$\pm$0.20 & 0.56$\pm$0.20 & 1.10$\pm$0.02 & 8.80$\pm$0.05 & 8.80$\pm$0.05 & 8.76$\pm$0.05 & 1.89$\pm$0.57 & 1.54$\pm$0.57 & 20.48$\pm$0.50 & 26.40$\pm$0.55  \\[0.5ex]
10 & 8.66$\pm$0.11 & 8.92$\pm$0.06 & -0.77$\pm$0.05 & -0.47$\pm$0.05 & -0.22$\pm$0.01 & 8.28$\pm$0.02 & 8.28$\pm$0.02 & 8.25$\pm$0.02 & 0.61$\pm$0.15 & 0.38$\pm$0.15 & 91.59$\pm$0.47 & 105.18$\pm$0.52  \\[0.5ex]
15 & 10.22$\pm$0.01 & 10.91$\pm$0.01 & -2.70$\pm$0.11 & -2.00$\pm$0.11 & -1.78$\pm$0.11 & 8.61$\pm$0.09 & 8.61$\pm$0.09 & 8.58$\pm$0.09 & 0.00$\pm$nan & 0.00$\pm$nan & 0.20$\pm$0.05 & 0.25$\pm$0.05  \\[0.5ex]
30 & 10.50$\pm$0.08 & 10.75$\pm$0.17 & 1.45$\pm$0.06 & 1.68$\pm$0.06 & 1.80$\pm$0.01 & 8.72$\pm$0.01 & 8.72$\pm$0.01 & 8.69$\pm$0.01 & 1.87$\pm$0.18 & 1.74$\pm$0.18 & 101.38$\pm$0.61 & 107.66$\pm$0.67  \\
\end{tabular}
\end{table*}

\end{appendix}
\end{document}